\documentclass[twocolumn]{aastex6} 

\usepackage{xspace}
\usepackage{graphicx}
\usepackage{natbib}
\usepackage{amsmath}

\newcommand{\nraoblurb}{The National Radio Astronomy Observatory is
a facility of the National Science Foundation operated under cooperative
agreement by Associated Universities, Inc.}
\newcommand{\hide}[1]{}

\newcommand{\gl}{\ensuremath{\ell}\xspace}
\newcommand{\gb}{\ensuremath{{\it b}}\xspace}

\newcommand{\vlsr}{\ensuremath{V_{\rm LSR}}\xspace}
\newcommand{\lb}{\ensuremath{(\gl,\gb)}\xspace}

\newcommand{\ngc}[1]{NGC\thinspace #1}

\newcommand{\kms}{\ensuremath{\,{\rm km\,s^{-1}}}\xspace}

\newcommand{\pc}{\ensuremath{\,{\rm pc}}\xspace}
\newcommand{\kpc}{\ensuremath{\,{\rm kpc}}\xspace}
\newcommand{\K}{\ensuremath{\,{\rm K}}\xspace}
\newcommand{\mK}{\ensuremath{\,{\rm mK}}\xspace}

\newcommand{\mhz}{\ensuremath{\,{\rm MHz}}\xspace}
\newcommand{\ghz}{\ensuremath{\,{\rm GHz}}\xspace}

\newcommand{\degree}{\ensuremath{^\circ}\xspace}

\newcommand{\jy}{\ensuremath{\,{\rm Jy}}\xspace}

\newcommand{\hii}{{\rm H\,{\footnotesize II}}\xspace}

\slugcomment{Accepted for publication in ApJ, 5/9/2016}
\shorttitle{HII Region Ionization of the ISM}
\shortauthors{Luisi et al.}

\begin{document}

\title{HII Region Ionization of the Interstellar Medium:\\
A Case Study of NGC~7538}

\author{Matteo~Luisi\altaffilmark{1}, L.~D.~Anderson\altaffilmark{1,2},
  Dana~S.~Balser\altaffilmark{3},
  T.~M.~Bania\altaffilmark{4}, 
  Trey~V.~Wenger\altaffilmark{5}}

\altaffiltext{1}{Department of Physics and Astronomy, West Virginia University, Morgantown WV 26506, USA}
\altaffiltext{2}{Adjunct Astronomer at the National Radio
  Astronomy Observatory, P.O. Box 2, Green Bank WV 24944, USA}
\altaffiltext{3}{National Radio Astronomy Observatory, 520 Edgemont Road, Charlottesville VA 22903-2475, USA}
\altaffiltext{4}{Institute for Astrophysical Research, Department of Astronomy, Boston University, 725 Commonwealth Ave., Boston MA 02215, USA}
\altaffiltext{5}{Astronomy Department, University of Virginia, P.O. Box 3818, Charlottesville VA 22903-0818, USA}

\begin{abstract}

Using data from the Green Bank Telescope, we analyze the radio continuum (free-free) and radio recombination line (RRL) emission of the compact \hii\ region \ngc7538 (Sharpless 158).  We detect extended radio continuum and hydrogen RRL emission beyond the photodissociation region (PDR) toward the north and east, but a sharp decrease in emission toward the south and west.  This indicates that a non-uniform PDR morphology is affecting the amount of radiation ``leaking'' through the PDR.  The strongest carbon RRL emission is found in the western PDR that appears to be dense.  We compute a leaking fraction $f_R = 15 \pm 5$\,\% of the radio continuum emission measured in the plane of the sky which represents a lower limit when accounting for the three-dimensional geometry of the region.  We detect an average $^4\textnormal{He}^+/\textnormal{H}^+$ abundance ratio by number of $0.088 \pm 0.003$ inside the \hii\ region and a decrease in this ratio with increasing distance from the region beyond the PDR.  Using \textit{Herschel Space Observatory} data, we show that small dust temperature enhancements to the north and east of \ngc7538 coincide with extended radio emission, but that the dust temperature enhancements are mostly contained within a second PDR to the east. Unlike the giant \hii\ region W43, the radiation leaking from \ngc7538 seems to only affect the local ambient medium. This suggests that giant \hii\ regions may have a large effect in maintaining the ionization of the interstellar medium.
\end{abstract}

\keywords{\hii\ regions -- ISM: abundances -- ISM: bubbles -- ISM: individual objects (NGC 7538) -- photon-dominated region (PDR) -- radio lines: ISM}


\section{Introduction}

Despite the fact that they are relatively rare, O-type stars have a
large impact on the
interstellar medium (ISM) at both large and small spatial scales. Their ultra-violet (UV) photons propagate
through molecular clouds, dissociating molecules and ionizing the
gas. Due to their intense radiation fields, O stars are 
surrounded by \hii\ regions of ionized plasma
\citep[see][]{Hoglund1965,Mezger1967}.  At the interface between the
fully ionized \hii\ region and the neutral medium surrounding it there
is a photo-dissociation region (PDR).  The ionized gas within
\hii\ regions can be studied using radio recombination line (RRL) and
radio free-free continuum emission, which have the benefit of being
essentially free from the effects of extinction, whereas their PDRs can
be studied using numerous molecular or atomic transitions.

The ISM of galaxies like the Milky Way contains low-density
($\sim$0.1\,cm$^{-3}$) diffuse ionized gas known as the ``warm
interstellar medium'' (WIM), first proposed by \citet{Hoyle1963}. The
WIM is a major, widespread component of the ISM, with a scale height
of $\sim$1500\pc and temperatures between 6,000 and 10,000\K
\citep[][]{Reynolds1989,Haffner2009}. Optical emission line measurements have shown
that the WIM is in a lower ionization state and is ionized by a softer
radiation field compared to gas in \hii\ regions \citep{Madsen2006}. 
The ``extended low-density medium'' \citep[ELDM; see][]{Gottesman1970,Mezger1978} 
is occasionally cited as another diffuse ionized component of the ISM, 
with a smaller scale height of $\sim$100\pc and a density of 1 to
10\,cm$^{-3}$.  The distribution of the ELDM was
found to be correlated with the location of \hii\ regions
\citep{Alves2012}.

It is still not completely understood how the WIM maintains
its ionization \citep{Haffner2009}.  While the radiation from
supernovae can contribute, it cannot provide the total energy
required \citep{Hoopes2003}. The most likely source of ionizing
photons is O-type stars \citep{Domgoergen1994,Madsen2006}, but it is
unclear precisely how radiation from the luminous O-type stars is able to propagate
across the kiloparsec size-scales required given the distribution of
the WIM. One suggestion is that superbubbles created by supernovae and stellar
winds provide low-density regions for photons to traverse Galactic
distances \citep[see][]{Cox1974,Dove2000,Reynolds2001,Terebey2003,Dale2005}. Another 
possibility is the existence of a two-component \citep{Wood2000} or fractal
\citep{Ciardi2002} ISM with a sufficient number of low-density paths.

If the O-type stars within \hii\ regions are maintaining the
ionization of the WIM, their radiation must either escape through
their dense PDRs or their PDRs must be clumpy. In a clumpy PDR, photons 
could escape along  low-density pathways in some directions and heat 
the ambient dust.
For the Galactic \hii\ region RCW\,120, \citet{Anderson2010} found
dust temperature enhancements that correlate with locations where the
PDR shows discontinuities at $8.0$\,$\mu$m, which suggests that for this 
region radiation is leaking through such small-scale ($\sim$0.3\,pc) inhomogeneities in the 
PDR. This result was supported by CO observations by \citet{Anderson2015}, who
determined that these holes are spatially correlated with deficits in CO 
emission at distinct molecular
velocities.  They further noted that $\sim$5\,\% of the H$\alpha$ emission 
of RCW\,120 (as measured in the plane of the sky) is found directly outside
these holes and that overall RCW\,120 is leaking $\sim$$25 \pm
10$\,\% of its emission beyond its PDR. Extended radio continuum and 
RRL emission has also been observed by \citet{Kim2001} around ultracompact 
\hii\ regions. They argue that these extended emission envelopes are mostly 
due to ionizing radiation from the exciting star of the compact region.

In a survey of 117 \hii\ regions with multiple hydrogen RRL
velocities, \citet{Anderson2015b} found that most multiple-velocity
regions are clustered near large star-forming complexes in the inner
Galaxy.  They posited that the additional velocity components are caused
by diffuse ionized gas along the line of sight, perhaps due to photons
from these large complexes leaking into the ISM.  Observations in
\citet{Anderson2011} support this hypothesis, as they found that for
multiple-velocity \hii\ regions near W43, the strength of the diffuse
velocity component decreases with increasing projected distance from the central
position of W43.  This suggests that leaking photons from W43 may be a
major contributor to the WIM in these directions.  

The main goal of the present study is to understand how a single \hii\ region
may contribute to the diffuse ionized gas detected by \citet{Anderson2011}.
While the \citet{Anderson2015b} study suggests a link between
\hii\ regions and diffuse ionized gas, additional research on individual
regions is required. The region W43 is at the tip of the Galactic bar, 
along a very complicated sight line, so the relationship between the diffuse 
ionized gas and \hii\ regions is still not clear.

Numerous studies of external galaxies indicate 
that a significant amount of ionizing radiation may leak from \hii\ regions.  
Averaged over the entire \ngc157
galaxy, \citet{Zurita2002} found that 30\,\% of the emitted ionizing
photons escape from \hii\ regions based on their H$\alpha$
emission model. Though this escape fraction by itself may not be sufficiently large 
to account for the total luminosity of the diffuse ionized gas, they suggested 
that almost all ionizing radiation escapes from the highest luminosity \hii\ regions. As a result, the ionization 
of the diffuse gas could be maintained even with a much lower escape fraction 
of less luminous regions \citep[see][]{Zurita2000}. \citet{Oey1997}
compared the estimated stellar Lyman continuum flux to the H$\alpha$ luminosities 
of the \hii\ regions for the Large Magellanic Cloud. They estimated that up to
50\,\% of ionizing radiation escapes the nebulae. \citet{Pellegrini2012} used
photoionization models of optically thin \hii\ regions. They calculated a 
lower limit on the ionizing escape fraction of 42\,\% for the Large Magellanic
Cloud and 40\,\% for the Small Magellanic Cloud. \citet{Giammanco2005} showed 
that models with optically thick clumps within the \hii\ region are 
generally in better agreement with observations than models that allow photon
transmission through optically thin clumps. They further found photon escape
fractions of 30\,\% -- 50\,\% for M51 and $<$60\,\% for M101.

The spectrum of the stellar radiation field outside an \hii\ region PDR
differs from that inside as some of the leaked photons undergo
absorption and re-emission processes in the local ISM. Though these
processes can be complex, a hardening in the H-ionizing continuum and a 
suppression of He-ionizing photons has
been observed \citep{Wood2004}. As there will be fewer photons with enough energy to ionize
He compared to H outside the \hii\ region, the observed ionic abundance ratio
$N(^4\textnormal{He}^+)/N(\textnormal{H}^+)$ should thus be lower than it
is inside \citep[see][]{Roshi2012}.  As well as photons being absorbed by the gas around
\hii\ regions, part of the radiation field is attenuated by
interstellar dust. This becomes particularly important in PDRs, as
they can absorb UV photons from the central source and re-emit them in
the infrared \citep{Hollenbach1997}. Knowledge of the dust properties
in and around the PDR can provide further understanding of the
radiative transfer effects in these regions
\citep[see][]{Compiegne2008}.

\begin{figure*}
\figurenum{1}
\centering
\begin{tabular}{rl}
\includegraphics[width=0.48\textwidth]{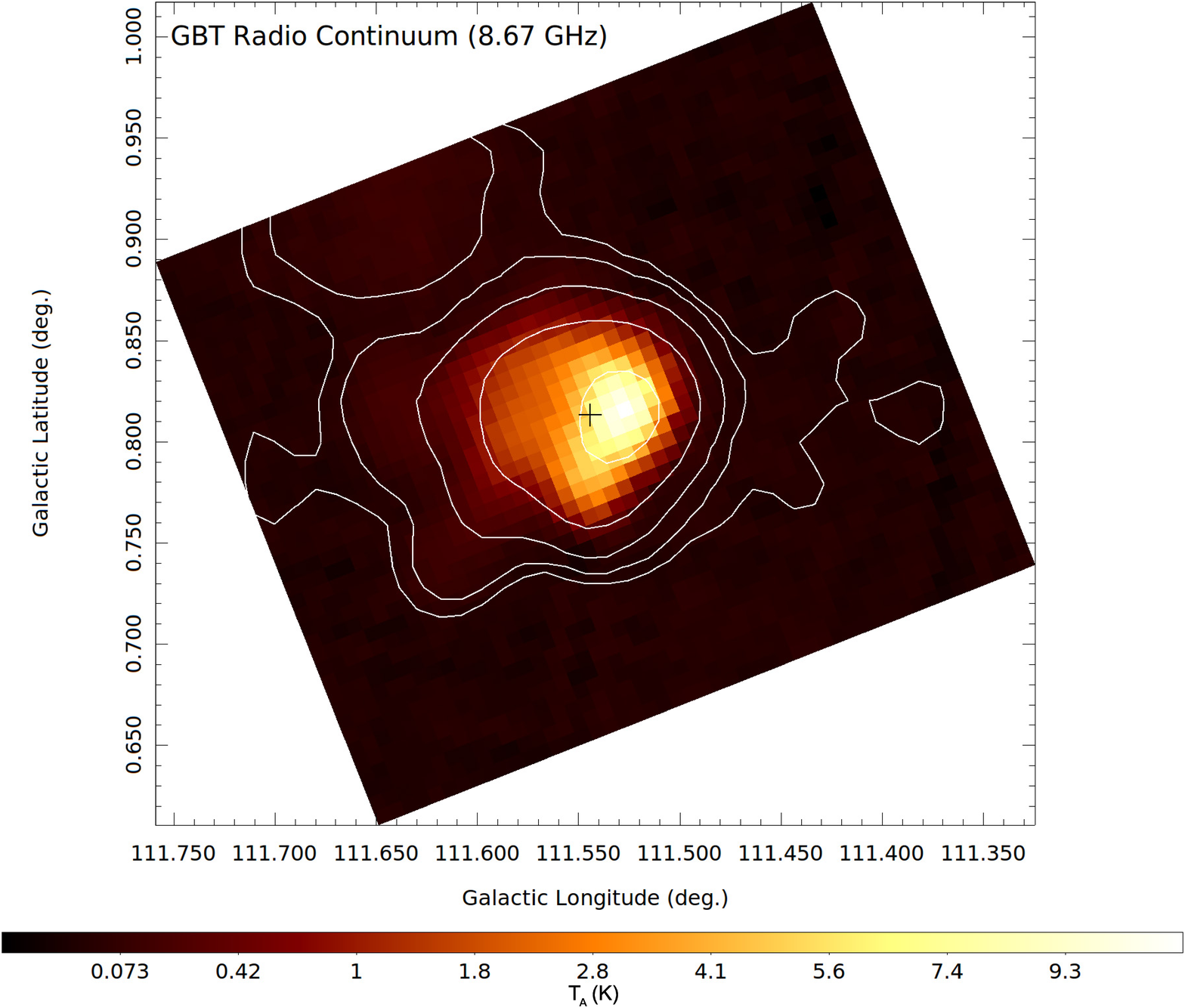}&
\includegraphics[width=0.48\textwidth]{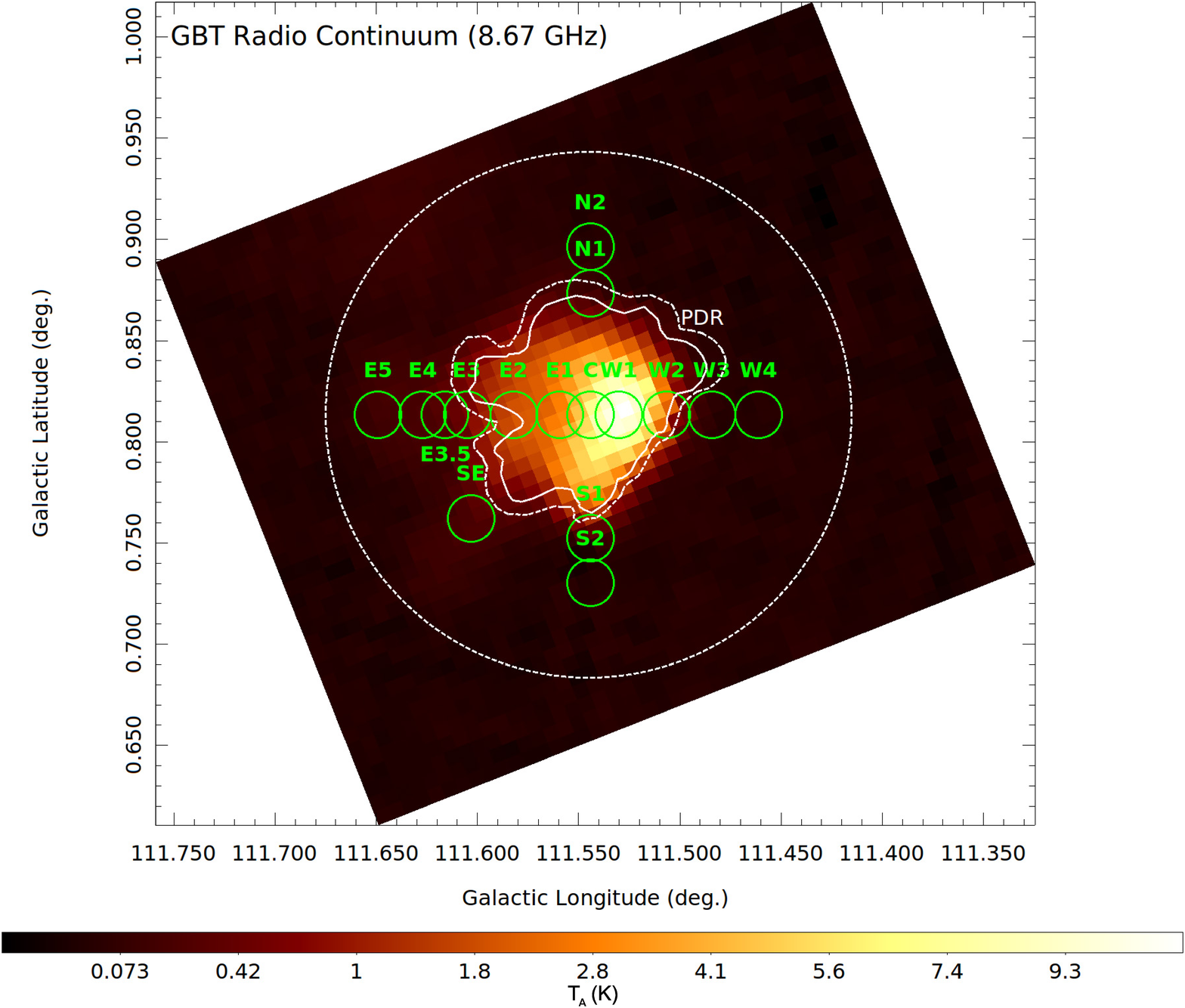}
\end{tabular}

\caption{Left: Radio continuum image of \ngc7538 at 8.67\ghz in units of
  antenna temperature. The contours start at $T_A=0.01\K$ and
  increase logarithmically to a maximum of $5\K$. The black cross marks 
  the central position (position ``C" in the right image) of the region.
  Right: Same image as left. The solid and dashed white regions 
  marked ``PDR" represent the inner and outer PDR boundary of
  \ngc7538 defined in Figure~\ref{fig:IRbeam} (see text).
  Green circles indicate the positions where RRLs were
  observed. The size of the circles is that of the GBT beam at 8.67\,GHz (82$''$). 
  Positions are labeled as in Table
  \ref{tab:RRLparam}. We use the
  area inside the large white dashed circle to derive the total
  intensity of the region. \label{fig:Loc}}
\end{figure*} \vspace{5pt}

Here we use radio observations of the \hii\ region \ngc7538 as a case study 
to understand the role that leaking UV radiation from \hii\ regions may have in
creating the diffuse ionized gas detected by \citet{Anderson2011}, and to better
understand the connection between this diffuse ionized gas in the vicinity of 
\hii\ regions and the WIM.  \ngc7538 is located in the Perseus spiral arm at a
distance of 2.65\kpc \citep[see][]{Moscadelli2009} and is
therefore close enough for detailed study.  It lies in the Outer
Galaxy where there is less confusion along the line of sight. 
We describe the radio continuum and RRL observations in Section~2 of this
paper. We derive an estimate of the location of the PDR boundary,
determine the percentage of leaking emission, and analyze the properties
of ionized gas outside the PDR in Section~3. We discuss these results
in Section~4 and conclude in Section~5.


\section{Observations}
We observed \ngc7538 in radio continuum and RRL emission at X-band
(9\,GHz; 3\,cm) using the National Radio Astronomy Observatory Green Bank
Telescope (NRAO GBT) from February 2013 to March 2014.  For both the
continuum and line data, we assume that the noise diodes, fired during
data acquisition, provide accurate intensities at the 10\,\% level. We
verified the flux density calibration of 3C147 in a
nearly concurrent program in November 2012 that used the same instrumental
configuration \citep{Anderson2015c} and found agreement with 
\citet{Peng2000} to within 10\,\%.  Throughout, we use a GBT X-band
gain of 2\,K\,Jy$^{-1}$ to convert from antenna temperature to flux density
\citep{Ghigo2001}.

\subsection{Radio Continuum}
We mapped the radio continuum emission from a 40$'$ square region
centered at ($\ell$,\,$b$) = (111.544\,\degree,\,0.813\,\degree) or
(J2000 R.A.,\,decl.) = (23:13:40,\,61:30:13) using the Digital
Continuum Receiver (DCR) on the GBT. Our central position for \ngc7538
corresponds to the 12\,$\mu$m 
Wide-Field Infrared Survey Explorer catalog 
\citep[WISE; see][]{Wright2010}. The
observations were done at 8665\mhz with a bandwidth of 320\mhz in two
orthogonal polarizations (left and right circular).  Radio continuum
emission from \hii\ regions at this frequency is caused by free-free
emission.  We slewed the telescope at a rate of 60\,$'$\,min$^{-1}$,
while sampling the total power every 100\,ms. We took data by scanning in Galactic
longitude and Galactic latitude, and created maps by averaging the two
directions and polarizations to minimize in-scan artifacts.

We show the radio continuum image in Figure~\ref{fig:Loc}.  
Here, all references to east or west refer to increasing or decreasing
Galactic Longitude, respectively, and references to north or south refer
to increasing or decreasing Galactic Latitude. Most of the
emission is peaked slightly west of the central location, 
but there is extended emission toward
the east that appears in two lobes.  There is also faint extended
emission toward the north, but little toward the south and west.

\begin{figure*}
\figurenum{2}
\centering
\begin{tabular}{cccc}
\includegraphics[width=.23\textwidth]{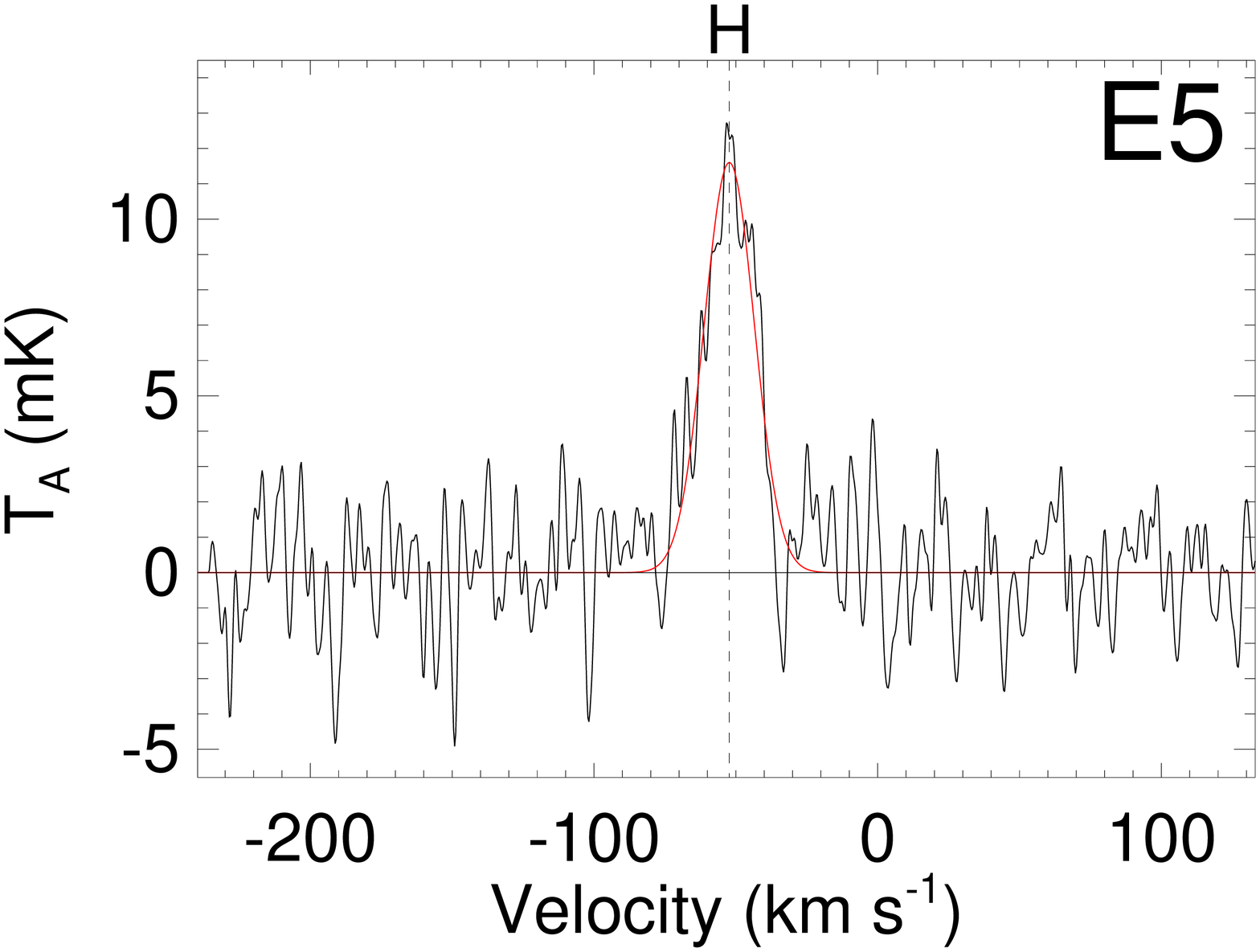} &
\includegraphics[width=.23\textwidth]{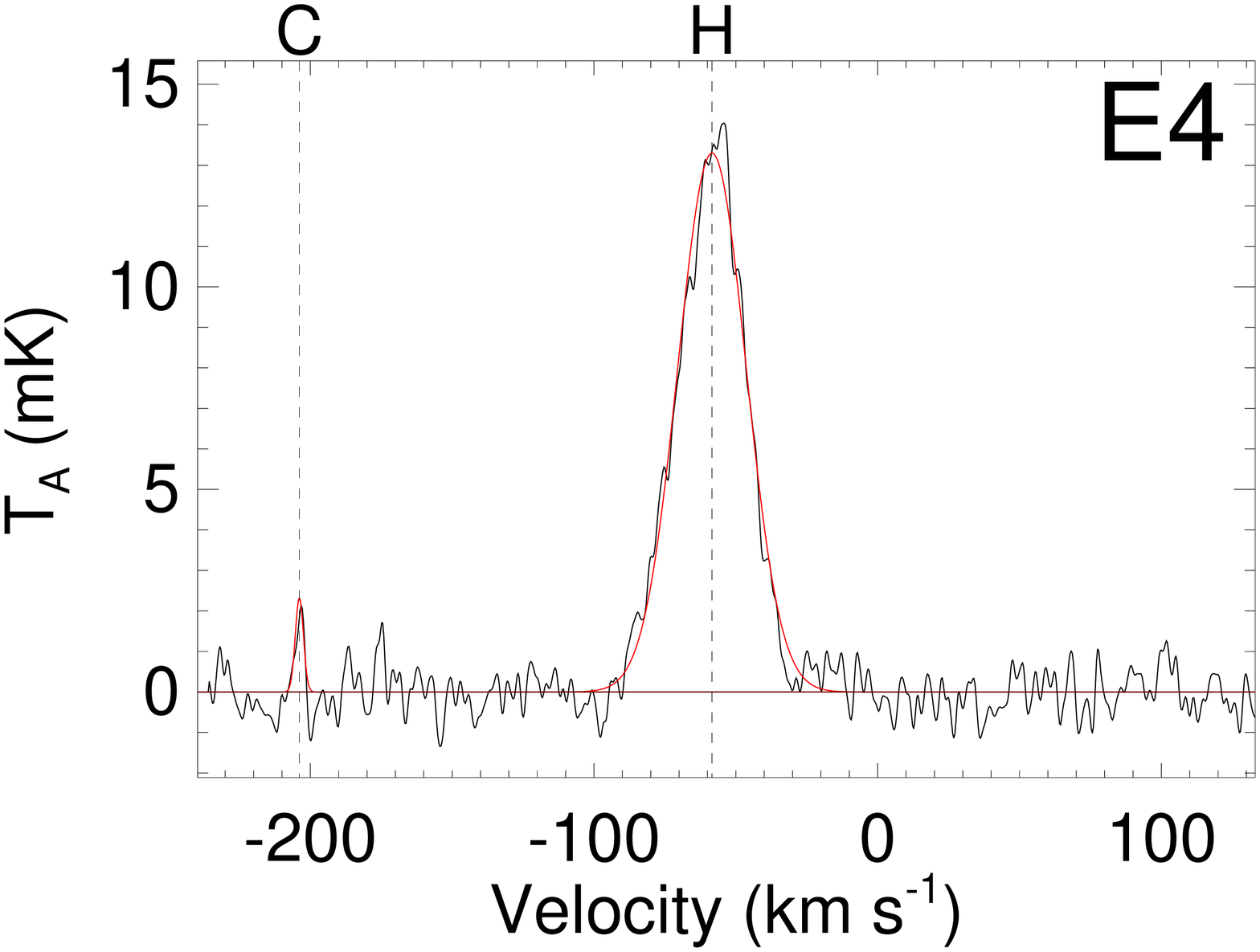} &
\includegraphics[width=.23\textwidth]{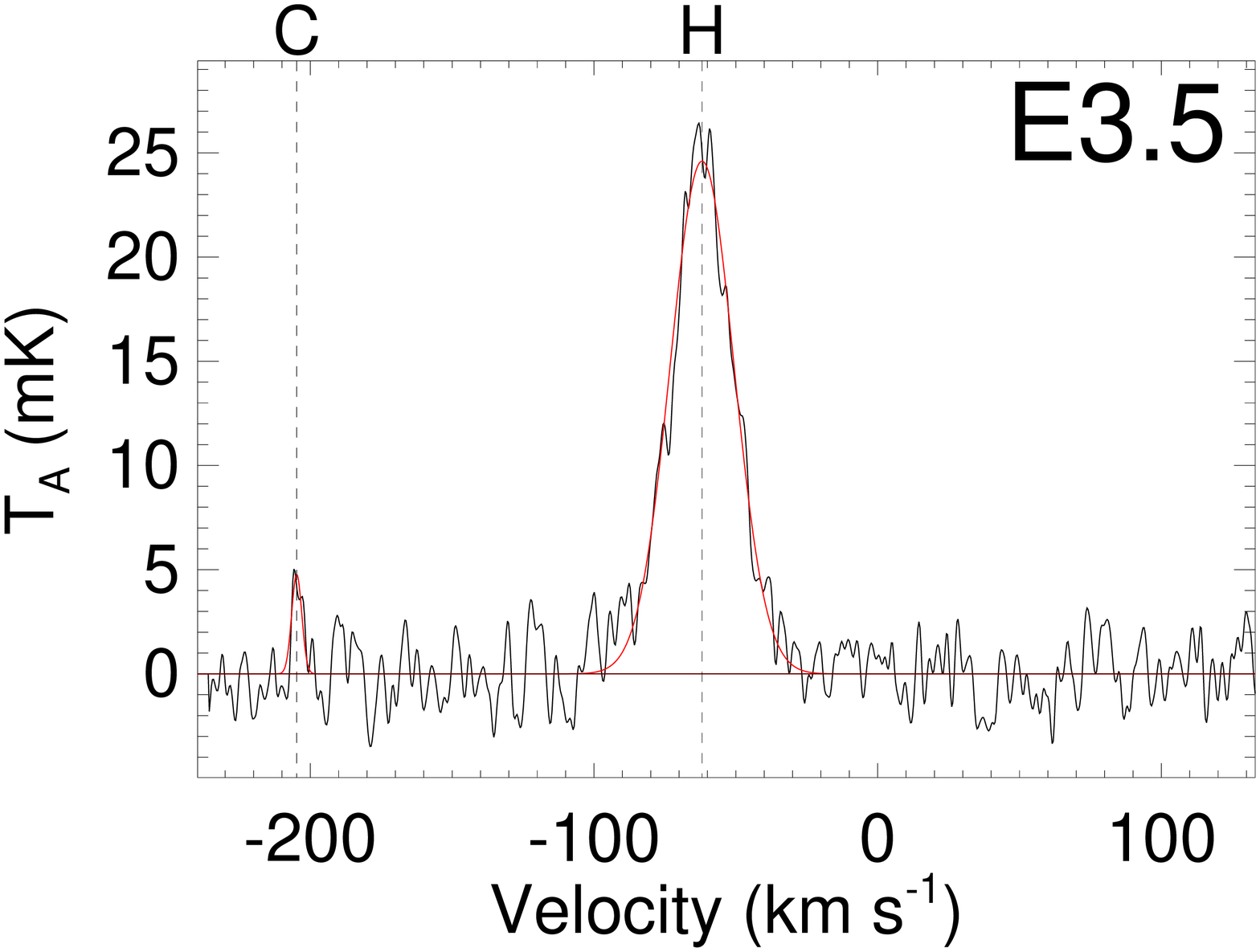} &
\includegraphics[width=.23\textwidth]{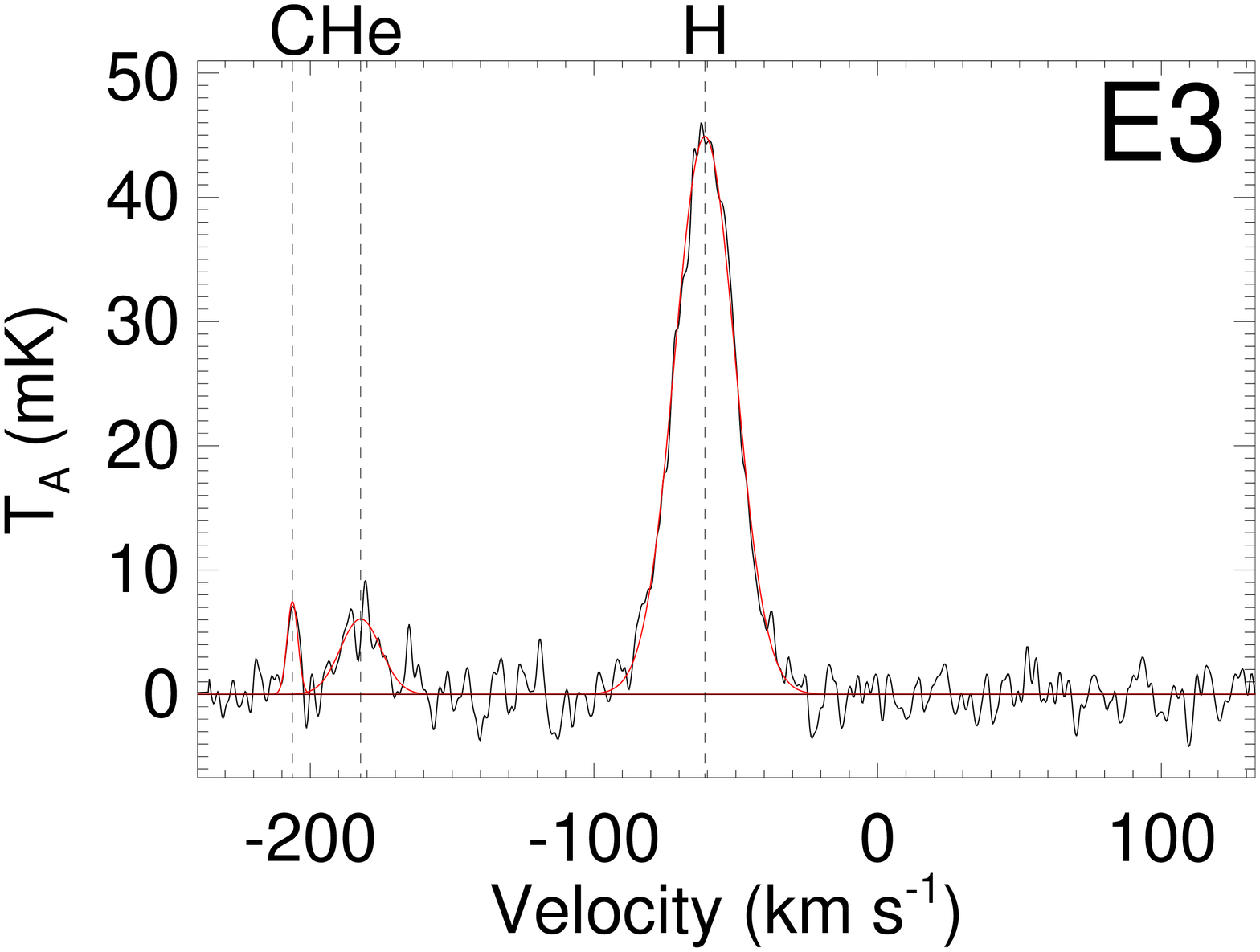} \\
\includegraphics[width=.23\textwidth]{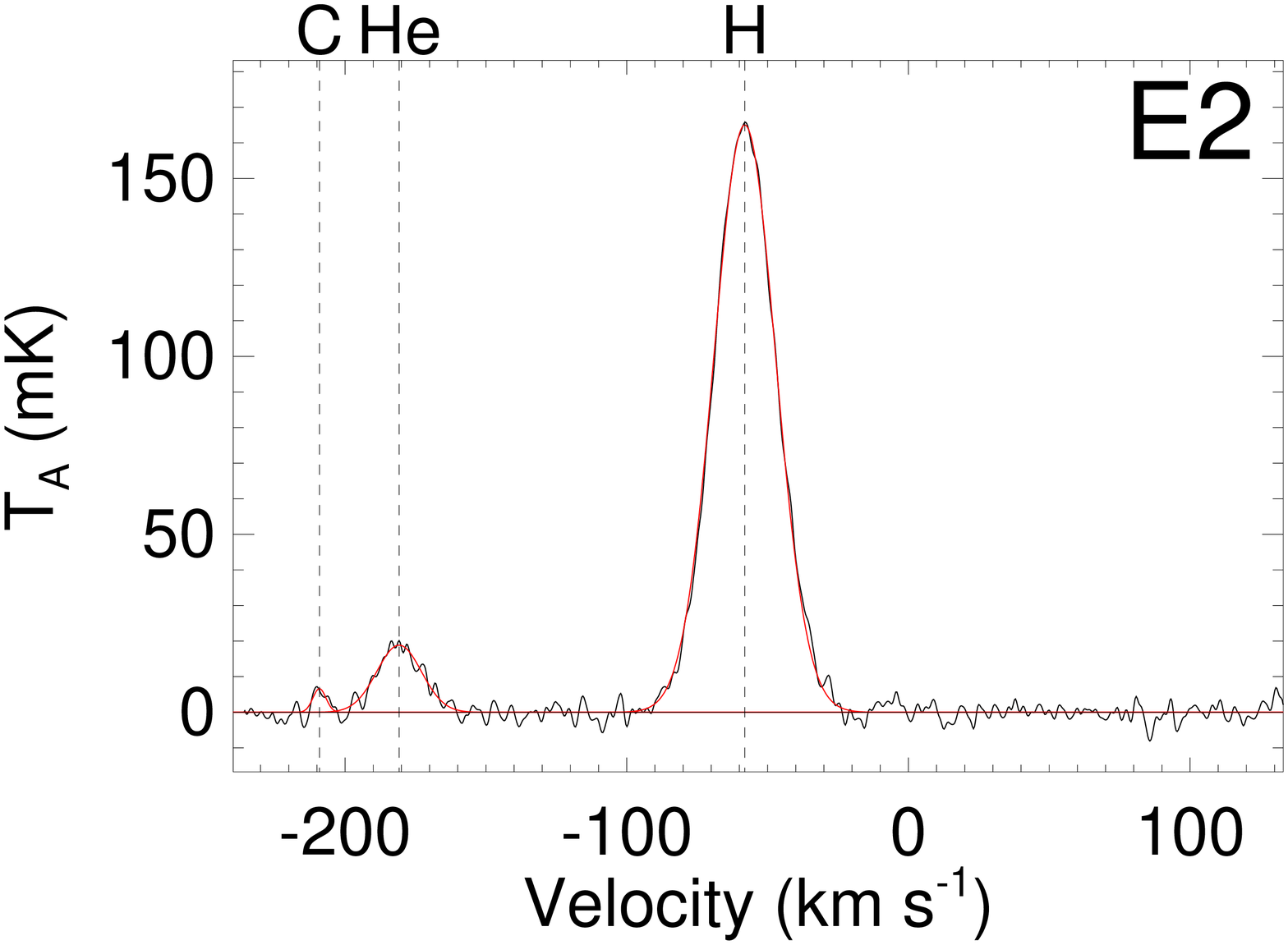} &
\includegraphics[width=.23\textwidth]{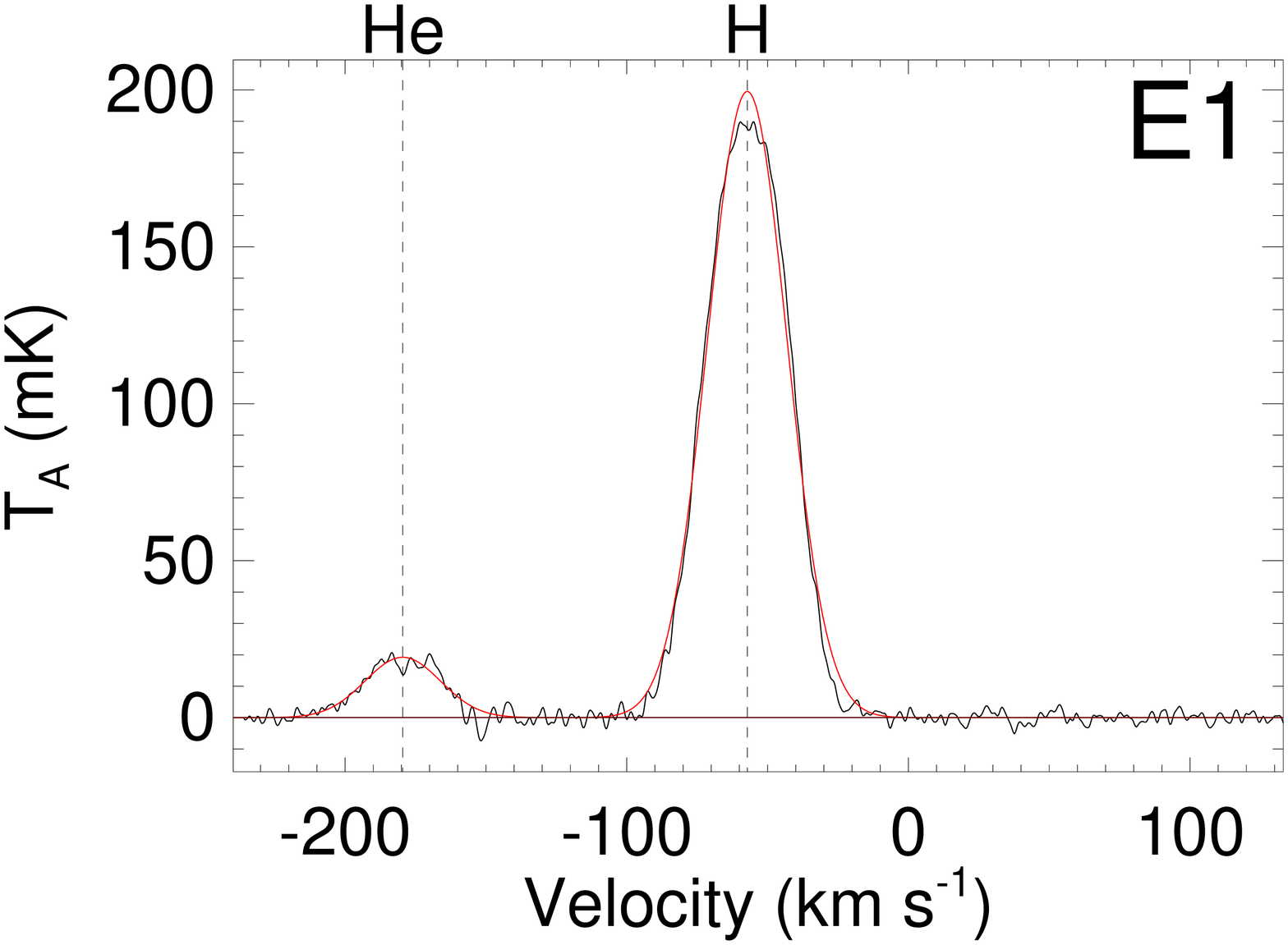} &
\includegraphics[width=.23\textwidth]{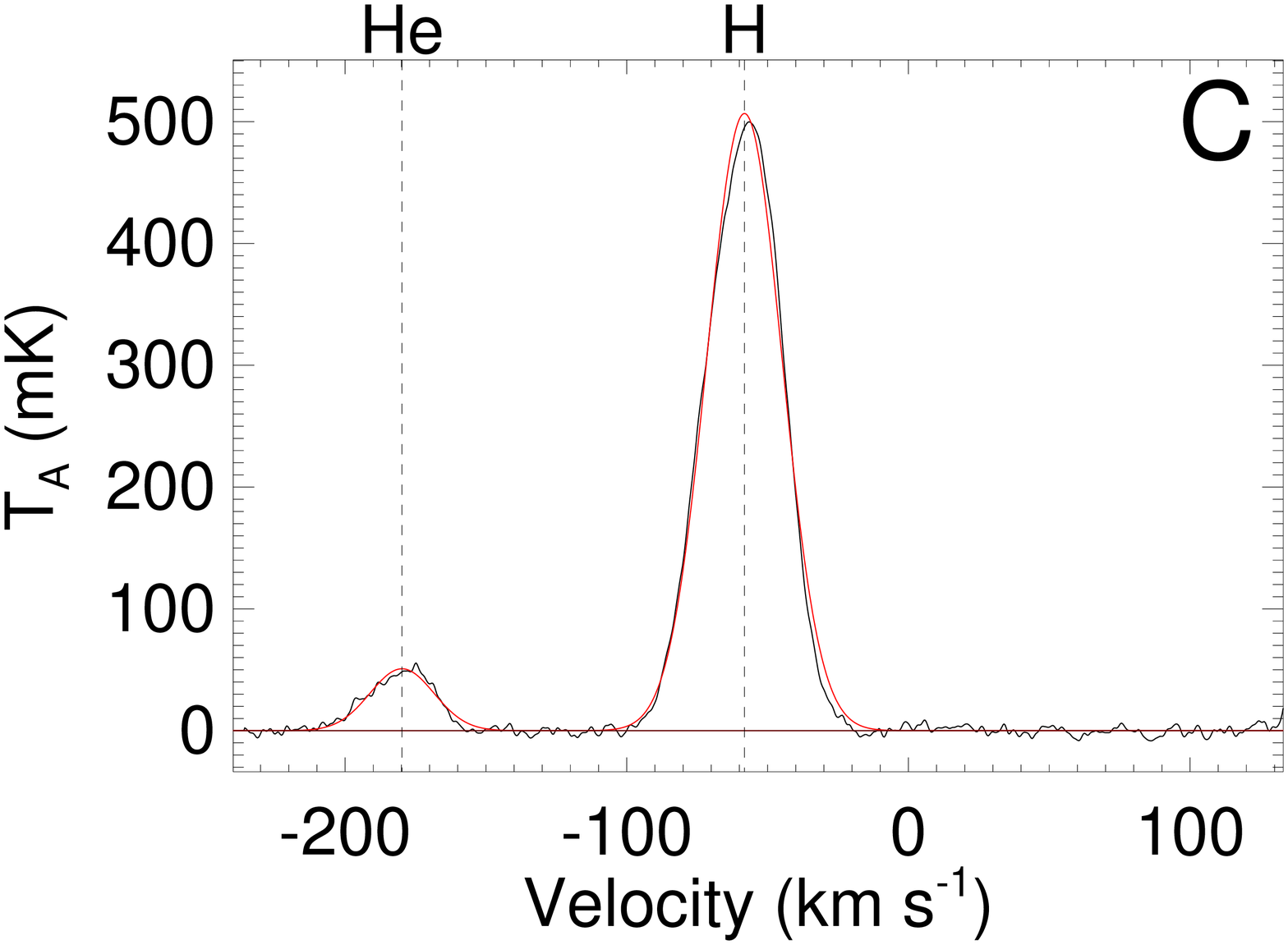} &
\includegraphics[width=.23\textwidth]{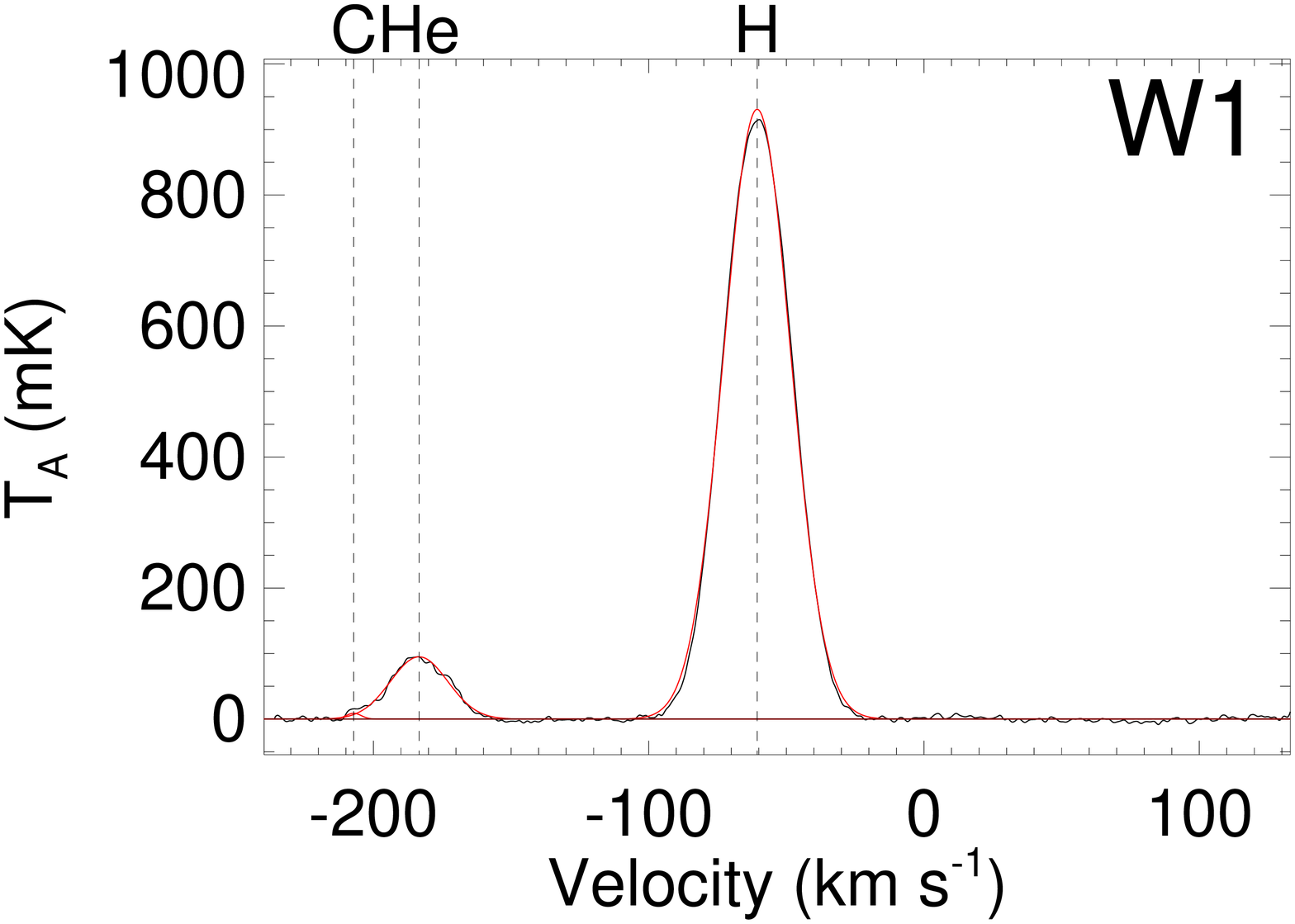} \\
\includegraphics[width=.23\textwidth]{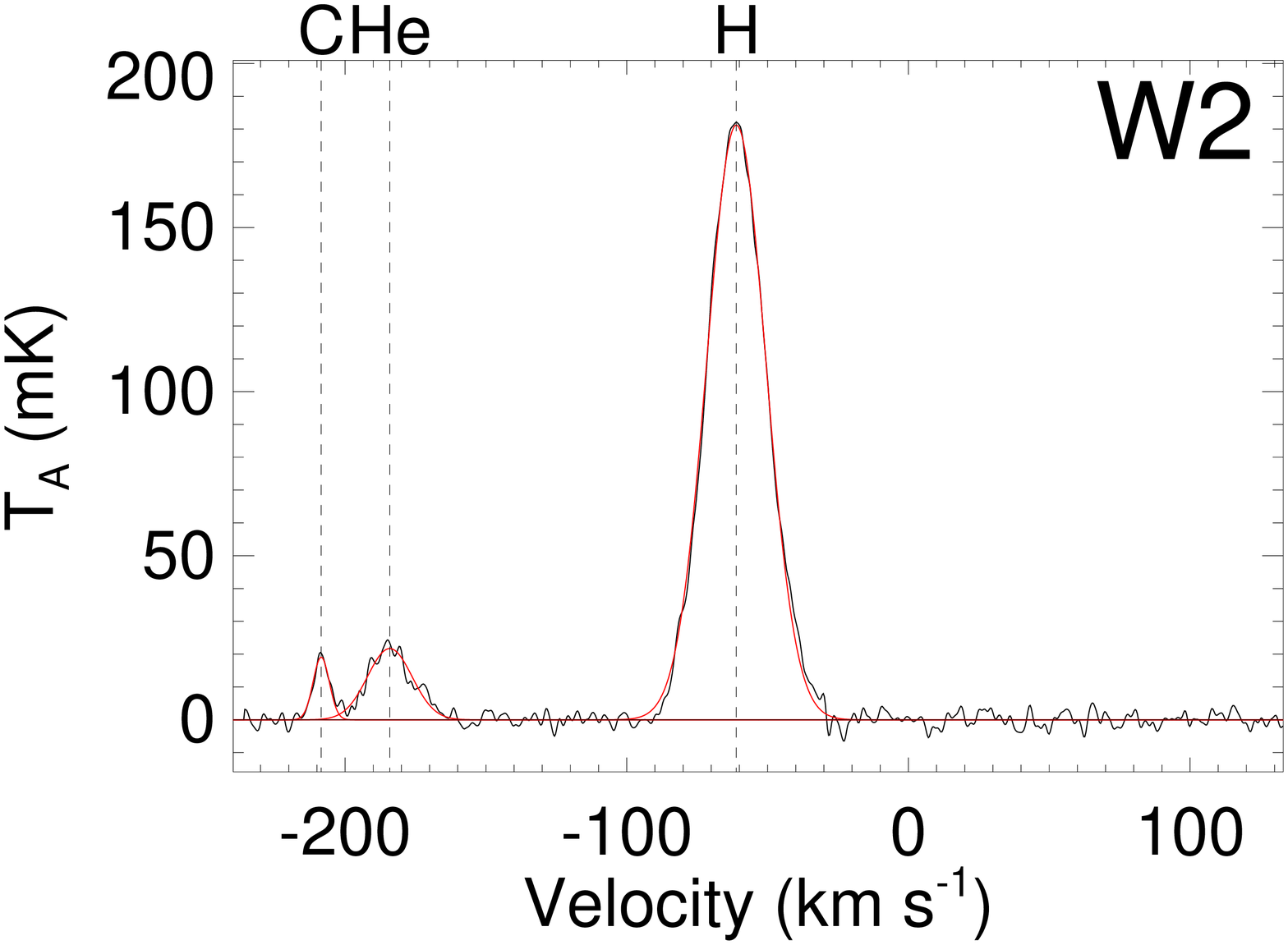} &
\includegraphics[width=.23\textwidth]{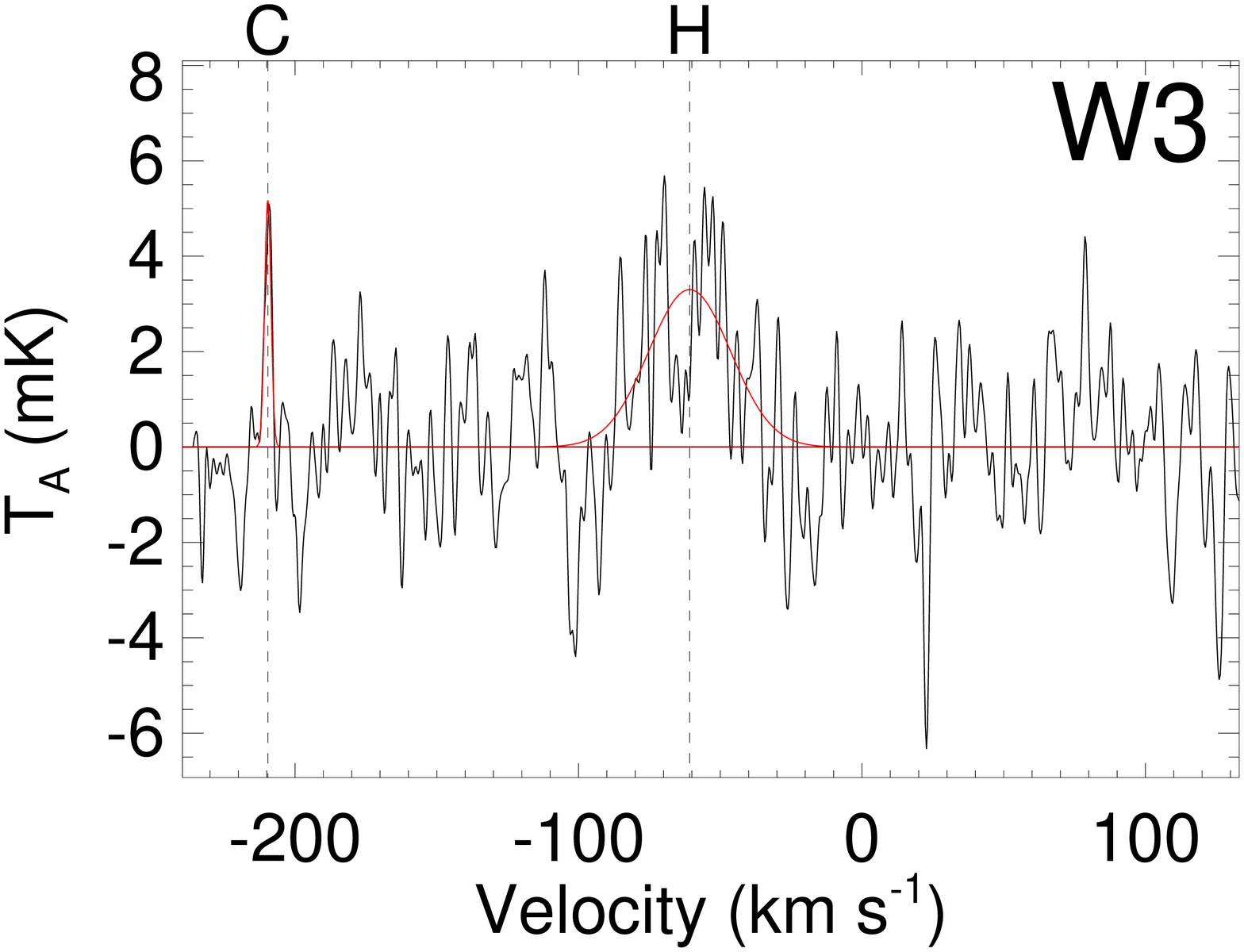} &
\includegraphics[width=.23\textwidth]{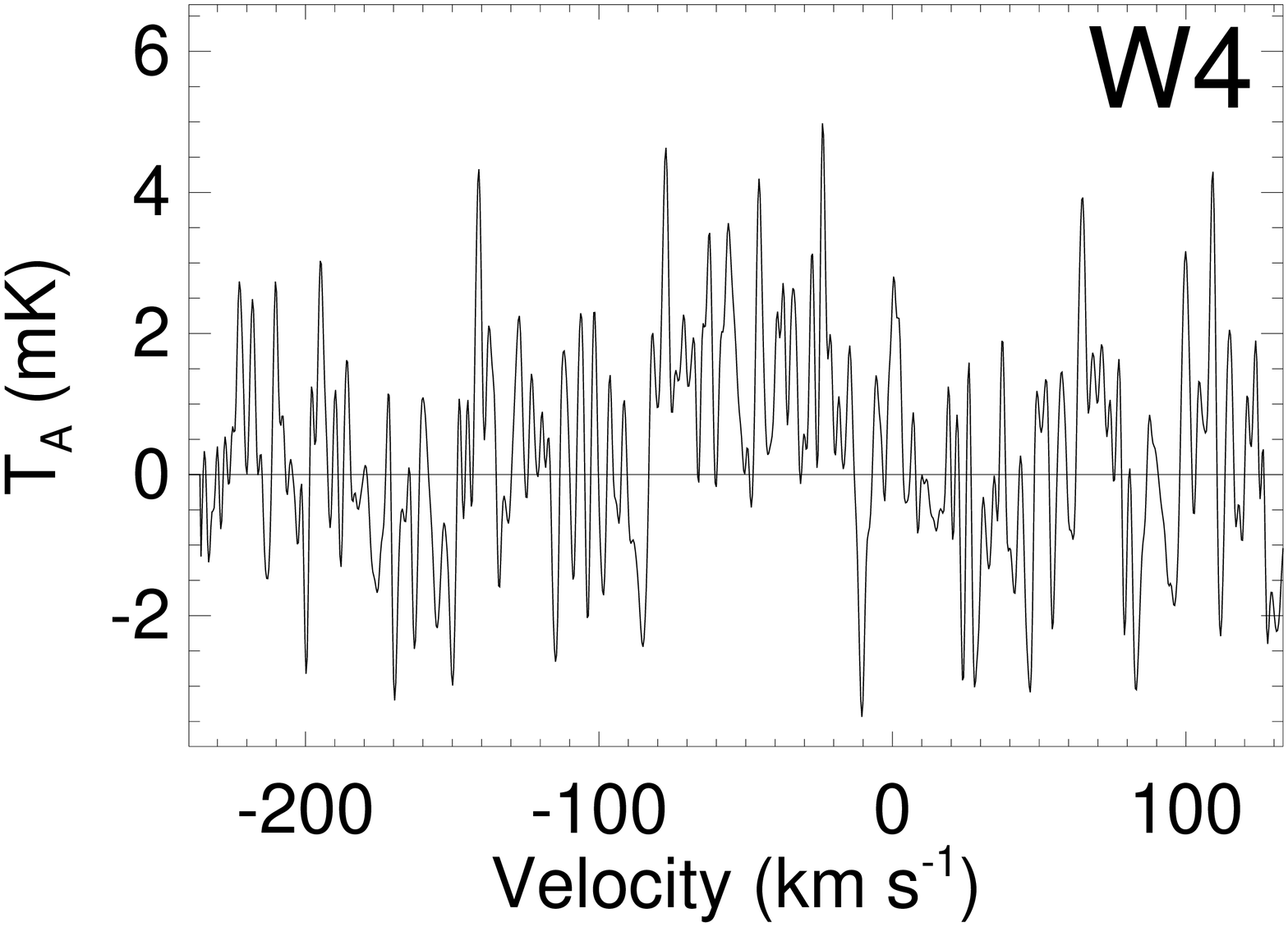} &
\includegraphics[width=.23\textwidth]{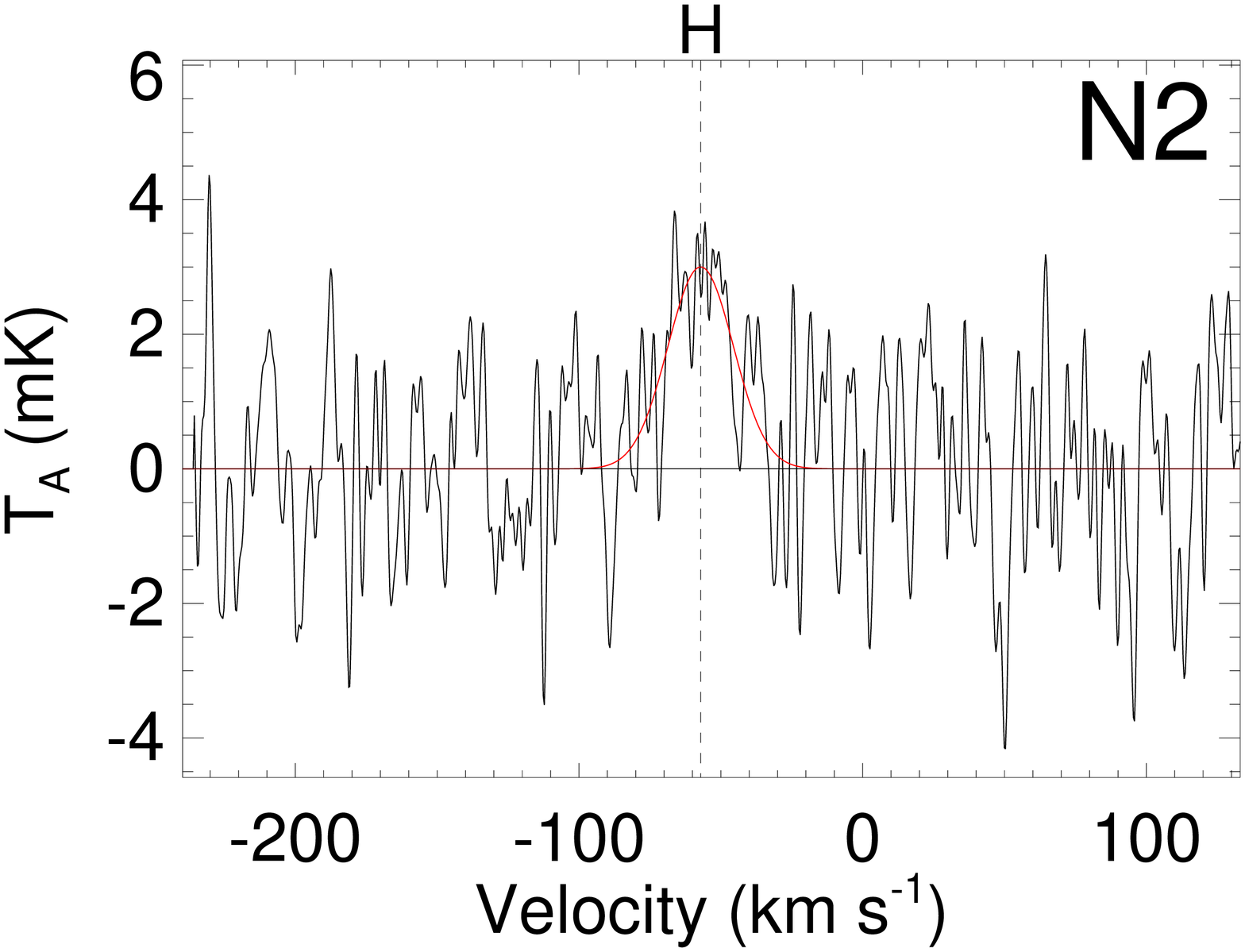} \\
\includegraphics[width=.23\textwidth]{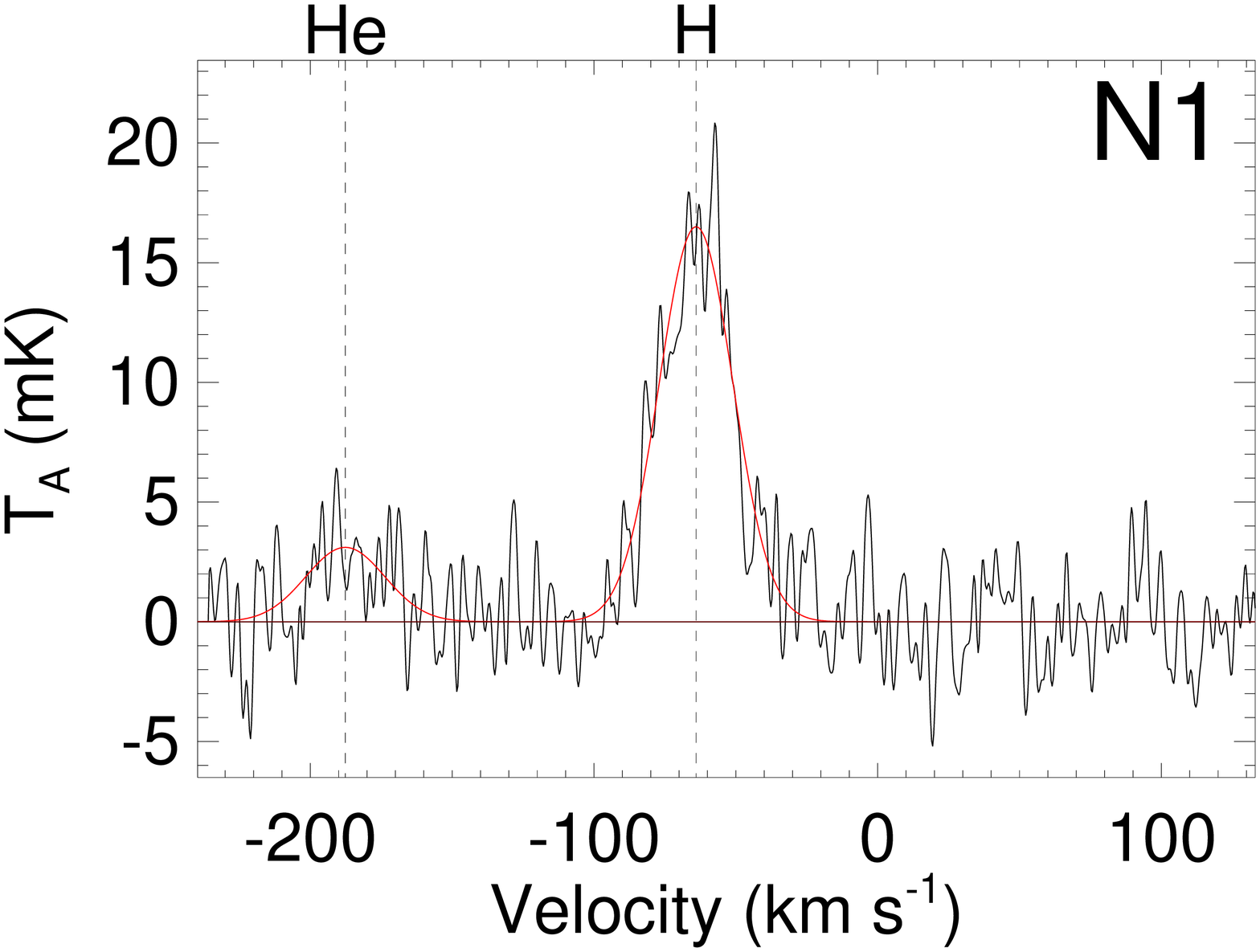} &
\includegraphics[width=.23\textwidth]{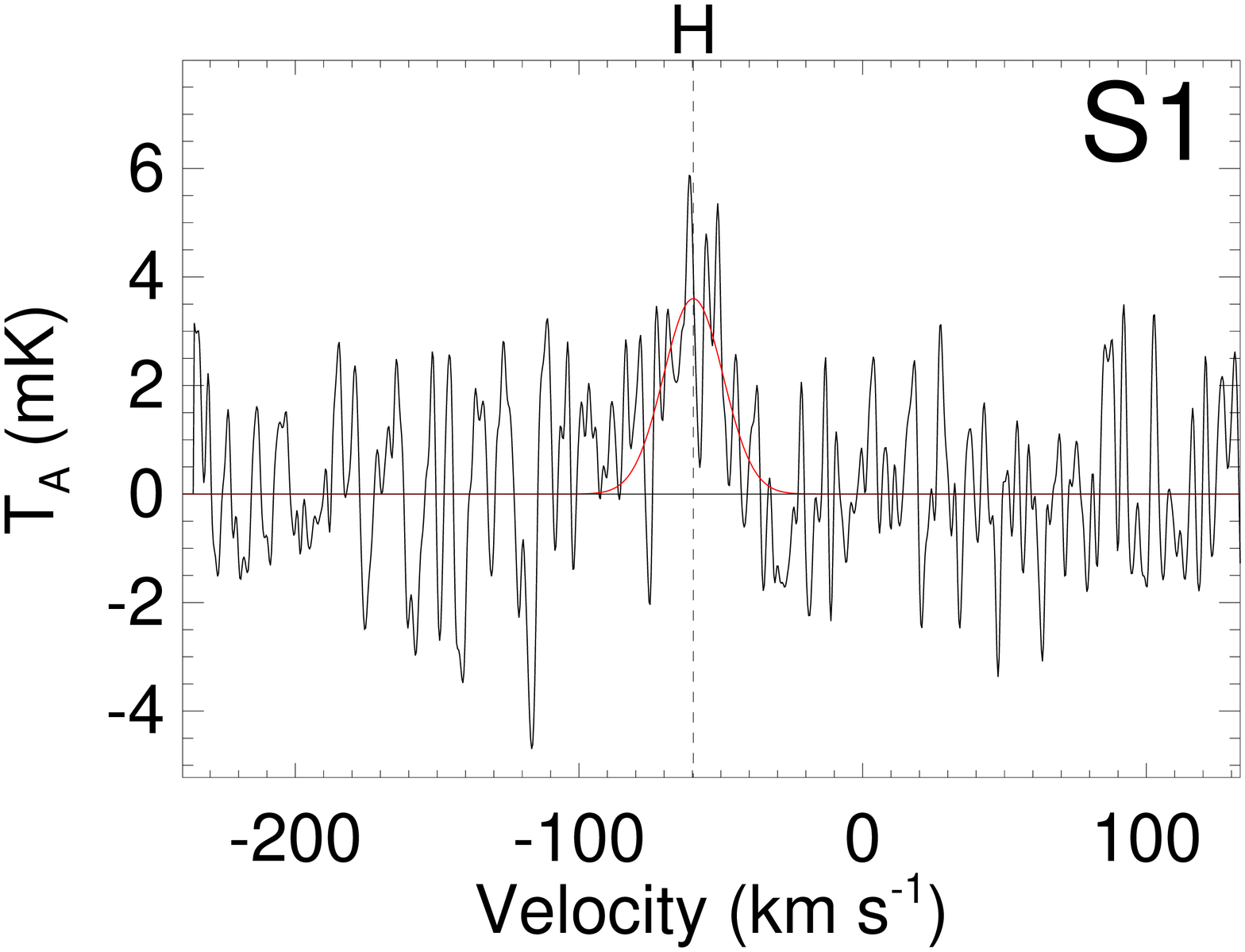} &
\includegraphics[width=.23\textwidth]{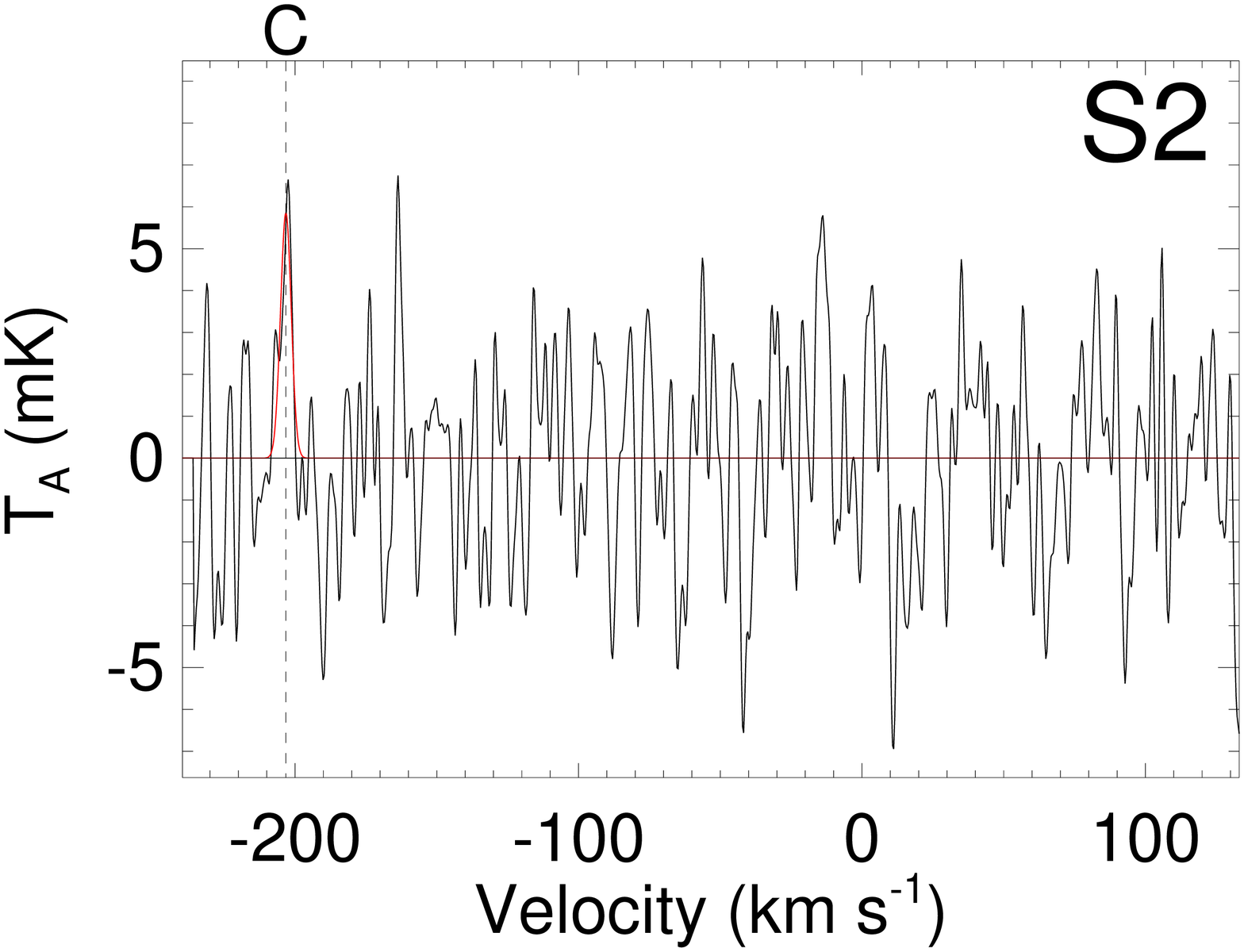} &
\includegraphics[width=.23\textwidth]{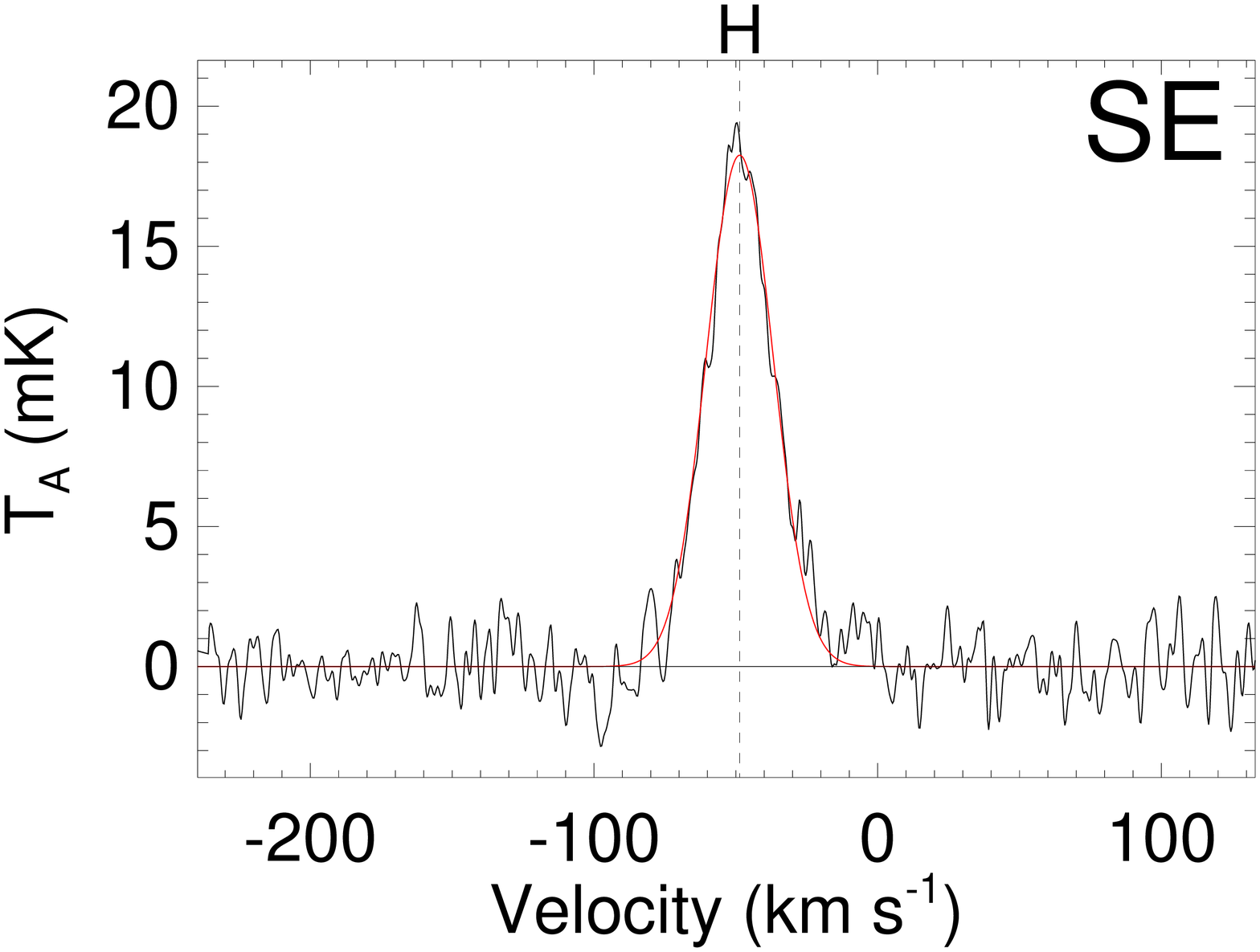} \\
\end{tabular}
\caption{RRL spectra of the 16 observed positions in \ngc7538, smoothed to a spectral
  resolution of 1.86\kms. Plotted is the antenna temperature as a function 
  of LSR velocity. Gaussian model fits to the H, He, and C peaks above the
  2$\sigma$ confidence interval are shown. The centers of the Gaussian
  peaks are indicated by dashed vertical lines. \label{fig:allspectra}}
\end{figure*}

\subsection{Radio Recombination Lines} \label{sec:RRLobs}
We made RRL observations
using the AutoCorrelation Spectrometer (ACS) on the GBT using our standard techniques \citep{Bania2010,Balser2011,Anderson2011}.  We employed
position switching with On\textendash{} and Off\textendash{}source
integrations of 6 minutes per scan. The Off\textendash{}source scans tracked the same azimuth and zenith angle path as the On\textendash{}source scans and were observed offset 6.5 minutes in right ascension. We simultaneously
measured 7 Hn$\alpha$ RRL transitions, H87$\alpha$ to
H93$\alpha$ (at rest frequencies of 9812.0, 9505.0, 9183.0,
8877.0, 8584.8, 8300.0, and 8045.6\mhz), in two orthogonal
polarizations.  Each observation therefore resulted in 14 independent
spectra.  Each spectrum spans 50\,\mhz and is sampled by 4096
channels, for a spectral resolution of $\sim$\,0.4\kms per channel.
The GBT HPBW at these frequencies ranges from 73$''$ to 90$''$. We
take the average of 82$''$ for all further RRL data analyses.

\begin{deluxetable*}{lccllrrrrrrrr}
\tabletypesize{\scriptsize}
\tablewidth{0pt}
\tablecaption{Radio Recombination Line Parameters \label{tab:RRLparam}}
\tablehead{
\colhead{Source}           & \colhead{$\ell$}      &
\colhead{$b$}          & \colhead{PDR\tablenotemark{a}} & \colhead{Line}  &
\colhead{$T_L$}          & \colhead{$\sigma \, T_L$}    &
\colhead{$\Delta V$}  & \colhead{$\sigma \, \Delta V$}  &
\colhead{$\vlsr$} & \colhead{$\sigma \, \vlsr$} &
\colhead{rms} & \colhead{$t_{\rm intg}$}\\
\colhead{}           & \colhead{(deg.)}      &
\colhead{(deg.)}      & \colhead{}    & \colhead{}  &
\colhead{(mK)}          & \colhead{(mK)}    &
\colhead{(km\,s$^{-1}$)}  & \colhead{(km\,s$^{-1}$)}  &
\colhead{(km\,s$^{-1}$)} & \colhead{(km\,s$^{-1}$)} &
\colhead{(mK)} & \colhead{(min)}}
\startdata
E5 & 111.649 & 0.813 & out & H & 11.6 & 0.3 & 20.8 & 0.6 & $-$52.3 & 0.3 & 1.6 & 24\\
E4 & 111.627 & 0.813 & out & H & 13.3 & 0.1 & 28.2 & 0.3 & $-$58.4 & 0.1 & 0.6 & 216\\
 & & & & C & 2.3 & 0.2 & 3.4 & 0.4 & $-$59.1 & 0.2 & & \\
E3.5 & 111.616 & 0.813 & out & H & 24.6 & 0.3 & 26.3 & 0.3 & $-$61.9 & 0.1 & 1.5 & 24\\
 & & & & C & 4.8 & 0.5 & 4.2 & 0.5 & $-$55.1 & 0.2 & & \\
E3 & 111.605 & 0.813 & out & H & 44.9 & 0.3 & 25.0 & 0.2 & $-$60.9 & 0.1 & 1.5 & 24\\
 & & & & He & 6.1 & 0.3 & 16.8 & 1.1 & $-$59.7 & 0.4 & & \\
 & & & & C & 7.5 & 0.6 & 4.5 & 0.4 & $-$56.5 & 0.2 & & \\
E2 & 111.582 & 0.813 & on & H & 165.0 & 0.4 & 25.8 & 0.1 & $-$58.1 & 0.1 & 2.4 & 12\\
 & & & & He & 18.8 & 0.4 & 18.7 & 0.4 & $-$58.2 & 0.2 & & \\
 & & & & C & 6.6 & 0.7 & 5.5 & 0.7 & $-$59.3 & 0.3 & & \\
E1 & 111.559 & 0.813 & in & H & 199.5 & 1.0 & 33.1 & 0.2 & $-$57.2 & 0.1 & 1.9 & 12\\
 & & & & He & 19.2 & 0.4 & 31.5 & 0.9 & $-$56.9 & 0.3 & & \\
W1 & 111.530 & 0.813 & in & H & 930.8 & 1.6 & 28.4 & 0.1 & $-$60.6 & 0.1 & 3.3 & 12\\
 & & & & He & 95.0 & 0.8 & 24.5 & 0.3 & $-$60.8 & 0.1 & & \\
 & & & & C & 8.9 & 0.5 & 6.4 & 0.5 & $-$57.4 & 0.2 & & \\
W2 & 111.507 & 0.813 & on & H & 181.2 & 0.6 & 24.6 & 0.1 & $-$61.1 & 0.1 & 2.0 & 12\\
 & & & & He & 21.7 & 0.4 & 18.6 & 0.4 & $-$61.5 & 0.2 & & \\
 & & & & C & 19.0 & 0.5 & 6.8 & 0.2 & $-$58.7 & 0.1 & & \\
W3 & 111.484 & 0.813 & out & H & 3.3 & 0.3 & 34.1 & 3.0 & $-$60.8 & 1.3 & 1.6 & 24\\
 & & & & C & 5.2 & 0.8 & 2.6 & 0.5 & $-$59.8 & 0.2 & & \\
W4 & 111.461 & 0.813 & out & \nodata & \nodata & \nodata & \nodata & \nodata & \nodata & \nodata & 1.4 & 24\\
N2 & 111.544 & 0.896 & out & H & 3.0 & 0.2 & 27.2 & 1.9 & $-$57.1 & 0.8 & 1.4 & 36\\
N1 & 111.544 & 0.873 & out & H & 16.5 & 0.4 & 30.5 & 0.9 & $-$64.0 & 0.3 & 0.9 & 12\\
 & & & & He & 3.1 & 0.3 & 32.8 & 4.6 & $-$65.1 & 1.6 & & \\
C & 111.544 & 0.813 & in & H & 506.7 & 2.2 & 31.3 & 0.2 & $-$58.2 & 0.1 & 3.3 & 12\\
 & & & & He & 50.8 & 0.8 & 26.7 & 0.5 & $-$57.2 & 0.2 & & \\
S1 & 111.544 & 0.752 & out & H & 3.6 & 0.2 & 24.7 & 1.9 & $-$59.7 & 0.8 & 1.5 & 24\\
S2 & 111.544 & 0.730 & out & C & 5.8 & 0.5 & 4.4 & 0.4 & $-$53.4 & 0.2 & 2.4 & 12\\
SE & 111.603 & 0.762 & out & H & 18.3 & 0.2 & 27.8 & 0.3 & $-$48.6 & 0.1 & 1.1 & 84
\enddata
\tablenotetext{a}{\,``out" = outside the PDR, ``on" = on the PDR, ``in" = inside the PDR.}
\tablecomments{All uncertainties are $\pm$1$\sigma$.}\vspace{-20pt}
\end{deluxetable*}

Figure~\ref{fig:Loc} shows our GBT 8.7\ghz radio continuum map of
\ngc7538 overlaid with the positions of the RRL measurements. We
observed a total of 16 positions, of which 15 are in a ``cross" 
with arms of constant latitude and
longitude to trace variations across the PDR boundaries. Eleven of the
observed positions are on a line of constant Galactic Latitude, which
we refer to as the East-West (EW) direction, and five positions are on
a line of constant Galactic Longitude. The cross directions intersect
at the central position at ($\ell$,\,$b$) = (111.544\,\degree,\,0.813\,\degree) or (J2000 R.A.,\,decl.) =
(23:13:39.8,\,61:30:13). 
We observed one additional position offset to
the south-east at ($\ell$,\,$b$) = (111.603\,\degree,\,0.762\,\degree)
or (J2000 R.A.,\,decl.) = (23:14:16.5,\,61:28:40).  Based on our definition of the location of the PDR boundary of \ngc7538 (see Section \ref{sec:PDR}), most of the
observed positions are outside the PDR, with only three
positions located within the PDR and two on the PDR.

We reduce and analyze the data in TMBIDL\footnote{V7.1, see
https://github.com/tvwenger/tmbidl.git.} \citep{Bania2014}.  For each position 
we average all 14 spectra together to increase the RRL
signal-to-noise ratio \citep[see][]{Balser2006} after first
re-gridding to the velocity resolution of the H87$\alpha$ data
(which has the poorest resolution) and shifting the spectra so they
are aligned in velocity.  When averaging, we use a weighting factor of
$t_{intg}/T^2_{sys}$, where $t_{intg}$ is the integration time and
$T_{sys}$ is the system temperature.  We smooth the resultant
spectrum to a resolution of 1.86\kms and fit a third-order polynomial baseline. We fit Gaussian models to the H, He, and C profiles if their peak signal was
two times the rms noise defined in a line-free portion of the
spectrum. We thus derive the peak line heights, their full width at half maximum (FWHM) values, and LSR
velocities for all detected H, He, and C components. In spectra with strong 
He lines, the carbon RRLs are often blended with He. This makes it difficult 
to fit both lines simultaneously. In these cases, we fit the He RRLs 
first, subtract the He Gaussian models, and then fit the C lines \citep[see][]{Quireza2006a}. In Figure~\ref{fig:allspectra}, we show the RRL data and Gaussian fits from all
observed positions.

\begin{figure*}
\figurenum{3}
\centering
\includegraphics[width=0.72\textwidth]{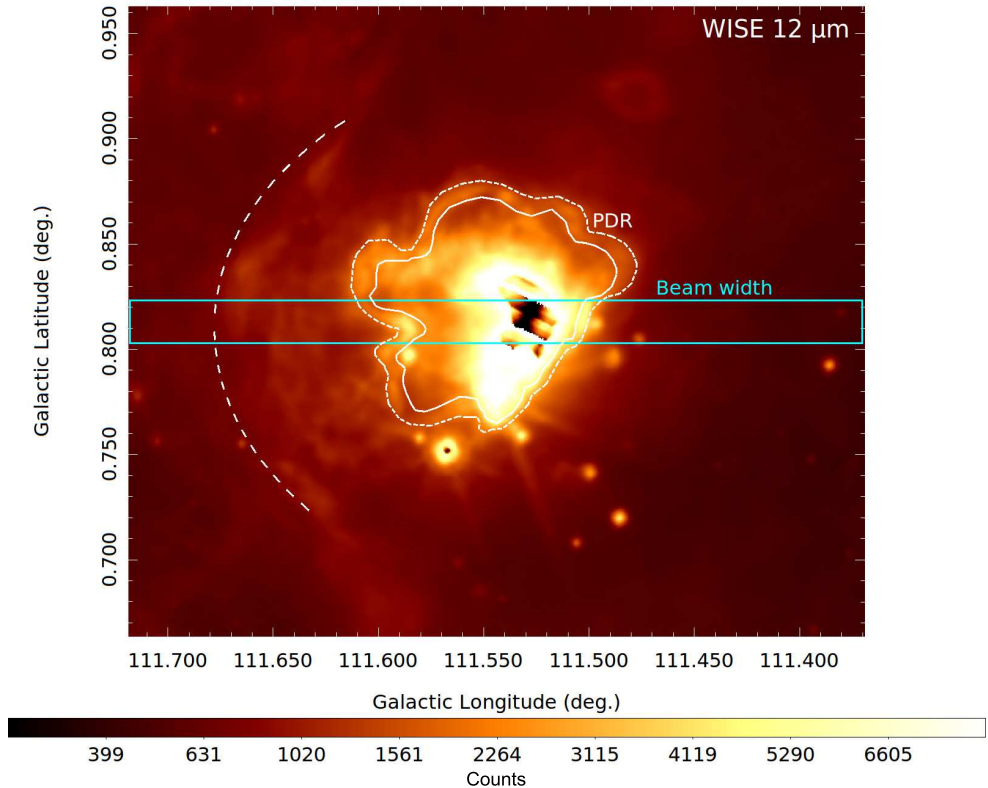}
\caption{WISE 12\,\micron\ image showing PAH emission from the
  region. The profile shown in Figure~\ref{fig:IRpdr} is extracted from the blue box area.  The width of the box is
  the GBT beam at 8.67\,\ghz\ and the vertical center of the box is at
  $b$ = 0.813\,\degree.  Our
  definitions of the inner and outer PDR boundaries are indicated by
  solid and dashed white lines marked ``PDR", respectively.  Another PDR boundary to the east is shown with the dashed white line.
  Portions of the interior of \ngc7538 are saturated in these data;
  we disregard these regions in the analysis. \label{fig:IRbeam}}
\end{figure*}

Of the 16 positions observed, we detect a hydrogen line at 
14 positions, a helium line at 7 positions, and a 
carbon line at 8 positions. We summarize the 
observations and RRL analysis in Table \ref{tab:RRLparam}, which lists
the source, the Galactic longitude and latitude, the location of the
observed position with respect to the PDR based on our PDR definition in
Section~\ref{sec:PDR} (``on'' for on the PDR, ``in'' for inside the 
PDR, ``out'' for outside the PDR), the line, the line intensity, the FWHM line width,
the LSR velocity, the rms noise in the spectrum, and the total integration 
time for each position, including all corresponding $1\sigma$ uncertainties
of the Gaussian fits.


\section{Ionized Gas in and around NGC\,7538}

\subsection{The PDR Boundary}
\label{sec:PDR}
If the origin of the WIM is linked to escaping ionizing
photons from \hii\ regions, we should detect a significant amount of
radio continuum emission ``leaking" through the PDR. In order to quantify this
amount, we need to define the inner PDR boundary. Radio emission
outside this boundary can then be classified as ``leaked", whereas
emission inside the boundary can be associated with the \hii\ region
itself.

We use the 12\,\micron\ WISE map shown in Figure~\ref{fig:IRbeam}
to define inner and outer PDR boundaries.  Within the
12\,\micron\ band is strong emission from a polycyclic aromatic
hydrocarbon \citep[PAH; e.g.][]{hollenbach97}.  This emission occurs
in PDRs of \hii\ regions, allowing us to use the image to estimate the
PDR extent.  Following the enhanced 12\,\micron\ intensity, we trace
by hand the inner and outer PDR boundaries around the region. While
the PDR to the north and east of the region is bright and
well-defined at 12\,\micron, the sharp decrease in intensity to the south-west makes
a clear distinction from the surrounding medium more difficult.

\begin{figure}
\figurenum{4}
\centering
\includegraphics[width=0.5\textwidth]{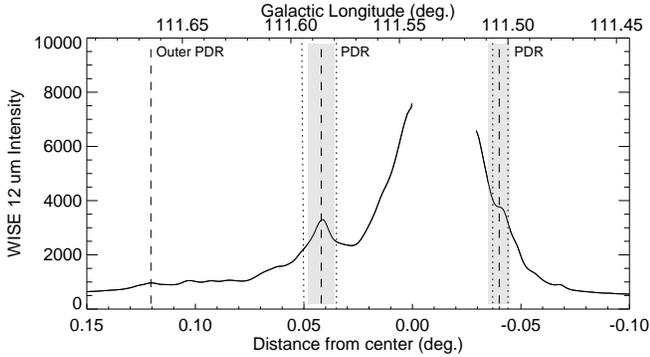}
\caption{Average of the WISE 12\,\micron\ intensity convolved with the GBT beam 
  for the boxed zone shown
  in Figure~\ref{fig:IRbeam}.  Missing data toward the center of
  the region were discarded due to saturation. We assume the center of
  the PDR is at the relative peaks in intensity at positions
  ($\ell$,\,$b$) = (111.586\,\degree,\,0.813\,\degree) and
  ($\ell$,\,$b$) = (111.504\,\degree,\,0.813\,\degree) as represented
  by the vertical dashed lines.  We estimate the FWHM width of the PDR
  using Gaussian fits to these data, as indicated by the shaded
  regions. The vertical dotted lines indicate the inner and outer PDR
  boundaries defined visually in Figure~\ref{fig:IRbeam}. The second, outer
  PDR to the east is shown as the dashed vertical line at $\sim$0.12\,\degree.
  \label{fig:IRpdr}}
\end{figure}

To test the plausibility of this PDR definition along the EW
direction (the line of $b$ = 0.813\,\degree), we convolve the WISE
12\,\micron\ data with a Gaussian of FWHM $87\arcsec$, the HPBW of the
GBT at the frequency of the radio continuum data
(Figure~\ref{fig:IRpdr}). This gives us a smoothed intensity distribution
of the 12\,\micron\ emission along the EW direction that can 
easily be compared to the radio data.
We assume that the center of the PDR can be
identified from peaks in the WISE intensity, which are located at
($\ell$,\,$b$) = (111.585\,\degree,\,0.813\,\degree) and
($\ell$,\,$b$) = (111.503\,\degree,\,0.813\,\degree), respectively. We
then fit a Gaussian to these peaks and define the width of the PDR as
the FWHM of the fits, resulting in a total width of 0.012\,\degree for
the eastern PDR boundary and 0.010\,\degree in the west. Due to
the convolution with the GBT HPBW these widths may be broadened. This
analysis agrees well with the visual definition (see Figure~\ref{fig:IRpdr}). PDRs, however, are characterized by a multitude of 
emission lines \citep{hollenbach97}. Using other emission lines to trace
the PDR may shift its location and therefore our characterization
of the PDR structure is by no means unique. 
Throughout the remainder of this work we use the visual
definition of the PDR boundary defined in Figure~\ref{fig:IRbeam}.

There is, however, a second PDR structure visible to the east in the
WISE 12\,\micron\ data of Figure~\ref{fig:IRbeam} (indicated by the
dashed white curve).  This second PDR may be an additional boundary
that prevents photons from escaping great distances from the region.
This second PDR is barely visible as a slight enhancement in
Figure~\ref{fig:IRpdr} at $\ell = +0.12$.

\subsection{The Leaking Emission Fraction}
\label{sec:leaking}
Since extinction due to interstellar dust is negligible at $\nu
\approx$ 8.7\ghz, we can use the radio continuum intensity to estimate
the percentage of emission escaping the inner PDR boundary of the region. 
To determine this percentage
of leaked emission, or escape fraction, $f_R$,
from the radio data the intensity inside and outside the PDR boundary
must be known. Assuming that the continuum emission is completely thermal 
and that the continuum background intensity is negligible, we sum the 
radio continuum map pixel values inside the visually-defined PDR
boundary to find the radio continuum intensity of the \hii\ region.
We then estimate the total intensity (inside and outside the PDR
boundary) by summing the pixel values from a circle of radius
0.13\,\degree centered at ($\ell$,\,$b$) =
(111.545\,\degree,\,0.813\,\degree) as shown in Figure~\ref{fig:Loc}. 
A faint source of radio continuum
emission is found to the north-eastern side of \ngc7538. 
We assume it is from a background radio source and
manually subtract its contribution to the intensity by excluding the source from the outside aperture. Since the source is faint and the continuum background intensity is insignificant, we assume that the uncertainty associated with the source subtraction is negligible.

Using a gain of 2\,K\,Jy$^{-1}$, we derive the total flux densities inside and
outside the PDR boundary of 17.3\,Jy and 3.0\,Jy, respectively, for 20.3\,Jy total.
Our flux density value for \ngc7538 itself (inside the PDR) is in
rough agreement with those found by other authors.
\citet{Gregory1991} found a value of 15.5\,Jy at 4.85\,GHz and
\citet{Becker1991} found 23.7\,Jy at 4.85\,\ghz.

\begin{figure*}
\figurenum{5}
\centering
\includegraphics[width=0.73\textwidth]{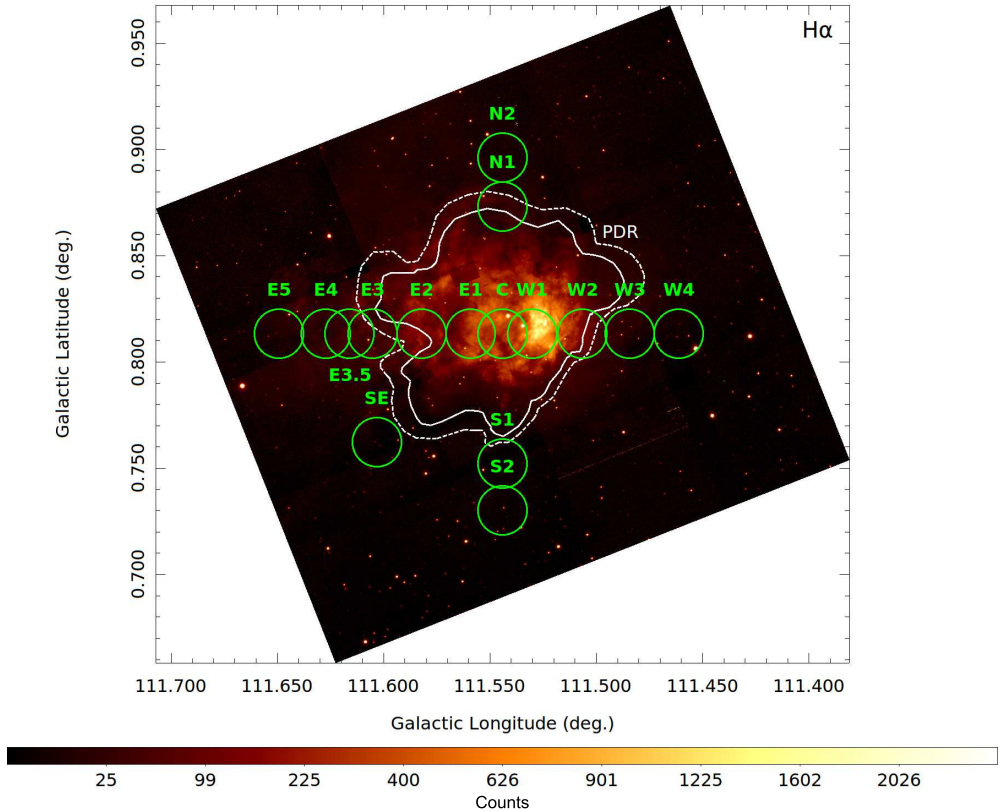}
\caption{H$\alpha$ emission image of \ngc7538. The solid and dashed
  white regions represent the inner and outer PDR boundaries as defined in Figure~\ref{fig:IRbeam}. As in
  Figure~\ref{fig:Loc}, green circles indicate the positions where
  RRLs were observed. \label{fig:Ha}}
\end{figure*}

We can calculate the expected flux density by making assumptions about 
the ionizing sources in \ngc7538. If two stars, O3V and O9V, are responsible 
for the observed emission \citep{Puga2010},
we estimate the number of emitted H-ionizing photons per second to be $N_{ly} = 10^{49.65}$
\citep[see][]{Martins2005}. From this, we calculate the expected flux density
using \citep[see][]{Rubin1968, Anderson2010a}
\begin{equation} 
N_{ly} = 4.76 \cdot 10^{48} \left( \frac{S_{\nu}}{\textnormal{Jy}} \right)
\left( \frac{T_e}{\textnormal{K}} \right) ^{-0.45} 
\left( \frac{\nu}{\textnormal{GHz}} \right) ^{0.1} 
\left( \frac{d}{\textnormal{kpc}} \right) ^2,
\end{equation}
where $N_{ly}$ is the number of H-ionizing photons per second, 
$S_{\nu}$ is the integrated
flux density at frequency $\nu$, $T_e$ is the electron temperature of the region
(see Section~\ref{sec:te}), and $d$ is the distance to \ngc7538. This yields
an expected flux density of 60\,Jy, about three times the measured value.
There is, however, debate about the spectral classification of the ionizing sources 
in \ngc7538. \citet{Puga2010} identify the main source of H-ionizing photons as a O3V star, whereas \citet{Ojha2004} classify it as a O5V star. Following the
classification of \citet{Ojha2004}, we would get an estimated $S_{\nu}$ of 
22.4\,Jy, much closer to the measured value of 20.3\,Jy. Moreover, part of the radiation field
may be absorbed by dust grains within the ionized region \citep[see][]{Arthur2004}
which would lower the observed flux density.

We calculate the escape fraction, $f_R = 15 \pm 5$\,\%, from the ratio of 
the outside to total (inside plus outside) radio continuum flux density using the inner PDR boundary from Figure~\ref{fig:Loc}. Using the outer PDR as our boundary, $f_R$ decreases to $\sim$8\,\%. 
We estimate the uncertainty in $f_R$ from two contributions.
We measure the effect of the uncertainty in the PDR location
by shifting the PDR by half its thickness and 
recomputing $f_R$. This gives a variation of $\sim$30\,\% in $f_R$,
which is the largest contribution to the total 
uncertainty. We also evaluate uncertainties
in the flux measurements due to the noise in the radio continuum map 
($\sim$10\,\% contribution in $f_R$). 
The uncertainty in $f_R$ takes into account both effects, added in 
quadrature.


As an alternative method, we measure the 656\,nm H$\alpha$ escape fraction 
of \ngc7538 using the Isaac Newton Telescope
Photometric H-Alpha Survey of the Northern Galactic Plane
(IPHAS) shown in Figure~\ref{fig:Ha}
\citep{Drew2005,Barentsen2014}. The H$\alpha$ IPHAS map has $\sim$80 times higher 
spatial resolution than the radio data.  
The disadvantages of using H$\alpha$ is that extinction
affects the results to an unknown degree and stellar subtraction in
regions of bright H$\alpha$ emission is difficult.  We determine the
leaking emission fraction, $f_{H \alpha}$, using the same methods as for
the radio continuum data (i.e., without performing stellar
subtraction) and find $f_{H \alpha} = 31 \pm 10$\,\%. The uncertainty
includes the same contributions used for the radio data. These, added in 
quadrature, result in an estimated uncertainty of $\sim$5\,\%.
Since the southern region of \ngc7538 shows less emission in H$\alpha$ inside 
the PDR compared to the radio map, we assume the H$\alpha$ data is affected 
by extinction here, which would artificially increase $f_{H \alpha}$. To account 
for this, we add an additional 5\,\% uncertainty contribution estimated
by measuring variations in the H$\alpha$ to radio continuum intensity ratio.

Because of the rather large GBT beam compared to the size of the
region, some of the radio continuum emission that appears outside the
\hii\ region PDR may actually be from inside the PDR.  To investigate
the magnitude of this effect, we smooth the H$\alpha$ map with the
GBT's $\sim$87\,$''$ beam.  We find the same leaking fraction,
$f_{H \alpha, smooth} = 31 \pm 10$\,\%, indicating that the GBT beam
size has not significantly affected our determination of $f_R$.

Since \ngc7538 is observed in the plane of the sky, we
cannot readily distinguish emission beyond the PDR along the line of
sight from that within the PDR.  This effect would artificially
increase the amount of emission we associate with the \hii\ region
since ``leaked'' emission is superposed on top of the region itself.
As a result, $f_R$ and $f_{H \alpha}$ represent lower limits on the total 
leaking emission fraction.

\begin{figure}
\figurenum{6}
\centering
\includegraphics[width=0.5\textwidth]{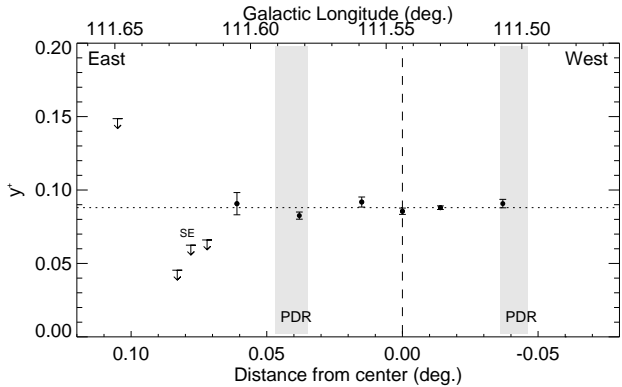}
\caption{The ionic abundance ratio, $y^+$, as a function of distance from the central
  location. The dotted horizontal line shows the average $y^+ = 0.088$ inside the PDR. All observed RRLs, with the exception of
  position SE (marked in the figure), were taken along the EW direction at a Galactic
  latitude of $b$ = 0.813\,\degree. The x-axis and meaning
  of the shaded regions are the same as in Figure~\ref{fig:IRpdr}. Error bars are $\pm$1$\sigma$.
  These measurements are consistent with a softening of the stellar radiation field
  outside the PDR. \label{fig:HeH}}
\end{figure}

\subsection{Ionic abundance ratio}
We use our RRL measurements to characterize the state of the ionized
gas inside and outside the PDR. Previous observational \citep{Hoopes2003} 
and theoretical work \citep{Wood2004} shows that radiation escaping the
\hii\ region has a softer spectrum due to absorption and re-emission
processes in the surrounding gas. This would indicate a decrease in
the $y^+ = N(^4\textnormal{He}^+)/N(\textnormal{H}^+)$ ionic
abundance ratio by number outside the PDR. We derive $y^+$ using
\begin{equation} y^+ = \frac{T_L(^4\textnormal{He}^+)\Delta V (^4\textnormal{He}^+)}{T_L(\textnormal{H}^+)\Delta V (\textnormal{H}^+)},\label{eq:heh} \end{equation}
where $T_L(^4\textnormal{He}^+)$ and $T_L(\textnormal{H}^+)$ are the
line strengths of He and H, respectively, and $\Delta V
(^4\textnormal{He}^+)$ and $\Delta V (\textnormal{H}^+)$ are the
corresponding FWHM line widths \citep{Peimbert1992}. If He was not detected, 
we use upper limits of $T_L(^4\textnormal{He}^+) = 3 \times \textnormal{rms}$ 
and $\Delta V (^4\textnormal{He}^+) = 0.84\, \Delta V (\textnormal{H}^+)$ 
to find $y^+$. Here, the constant $0.84$ is the average line width ratio $\Delta V
(^4\textnormal{He}^+) / \Delta V (\textnormal{H}^+)$ from our RRL data. This value is in agreement with \citet{Wenger2013}, who found $\Delta V(^4\textnormal{He}^+) / \Delta V (\textnormal{H}^+) = 0.77 \pm 0.25$, averaged over 54 individual \hii\ regions.

Results for the ionic abundance ratio, $y^+$, electron temperatures 
(see Section \ref{sec:te}), and
dust temperatures (see Section \ref{sec:dust}) are summarized in
Table~\ref{tab:calc}. Listed are the Galactic
longitude and latitude, the location of the observation with respect to the 
PDR (``on'' for on the PDR, ``in'' for inside the 
PDR, ``out'' for outside the PDR), the ionic abundance ratio, the LTE electron
temperature, and the estimated dust temperature (see next section).  All
uncertainties are $1\sigma$. In Figure~\ref{fig:HeH}, we show $y^+$ along the EW
direction. Though not located along the EW direction, we also include the
source SE at ($\ell$,\,$b$) = (111.603\,\degree,\,0.762\,\degree) in the 
plot. This position is surrounded by a faint lobe of radio
continuum emission outside the PDR similar to the eastern locations.

The ionic abundance ratio $y^+$ declines 
outside the PDR boundary but remains relatively constant inside. We find an 
average total ionic 
abundance ratio inside/on the PDR of $y^+ = 0.088 \pm 0.003$. Directly 
outside the eastern PDR boundary at a distance of $+0.061$\,\degree at 
position E3, $y^+$ is $0.091 \pm 0.008$, and then decreases 
with increasing distance. The large upper limit at a distance of 
$+0.105$\,\degree at position E5 is a result of the low signal-to-noise 
ratio of the observation and does not necessarily indicate an actual 
increase of $y^+$. We estimate an upper limit on the average $y^+ = 0.034$
for all locations with He non-detections outside the PDR.
The decline in $y^+$ with increasing distance 
suggests that the radiation field softens as the radiation passes through 
the PDR \citep{Reynolds1995,Hoopes2003,Wood2004,Roshi2012}. Since at 
position E3, which is directly outside our 12\,$\mu$m-defined PDR 
boundary, $y^+$ is comparable to that inside, it may indicate that 
the PDR extends slightly further to the east than our 
12\,$\mu$m-definition.

\subsection{Carbon RRLs}
\begin{figure}
\figurenum{7}
\centering
\vspace{-10pt}
\includegraphics[width=0.5\textwidth]{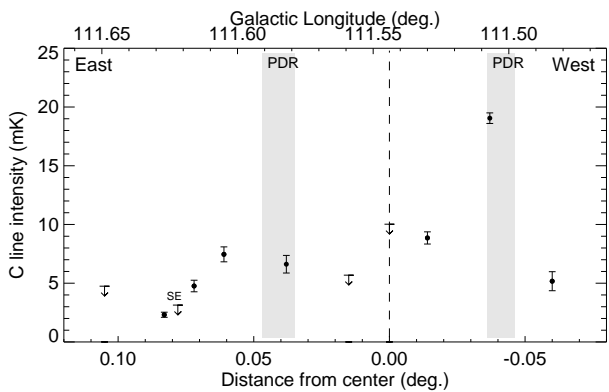}
\caption{Carbon line intensity as a function of distance from the central
  location in Galactic coordinates.  Shown are all C RRLs measured in the EW direction at a Galactic latitude of $b$ = 0.813\,\degree with the addition of position SE (marked). The x-axis and meaning
  of the shaded regions are the same as in Figure~\ref{fig:IRpdr}. Error bars
  are $\pm$1$\sigma$. \label{fig:CStrength}}
\end{figure}

Carbon RRL emission is primarily observed in PDRs \citep[see][]{Hollenbach1999}.
Because of its lower first ionization potential (11.2\,eV), carbon can be ionized 
by softer radiation than H (13.6\,eV) and He (24.6\,eV). This
softer radiation is better able to
pass through the \hii\ region and into the PDR surrounding it. Figure~\ref{fig:CStrength} shows the observed intensity of the carbon RRL,
$T_L(\textnormal{C})$, along the EW direction, again including the
position SE. If C was not detected, an upper limit defined by 3 times the 
observed rms is used. Here, we use the original spectral resolution of 1.86\kms as 
the typical C line profile is more narrow compared to H and He. As expected, 
we find the largest carbon RRL intensities near the PDR boundaries of the 
\hii\ region. 

The observed intensity of the line at position
W2 on the western PDR is $\sim$$19$\,mK, which is two times higher than that
detected at any other pointing.  It is probably no coincidence that
the W2 position falls on the PDR.  We hypothesize that this PDR is
dense and is absorbing the C-ionizing photons.  Position E2 on the
weaker eastern PDR also has a detected carbon line,
although only at an intensity of $\sim$$6$\,mK, comparable to values found 
both inside and outside the \hii\ region.

\begin{deluxetable*}{lcclrrrrrr}
\tabletypesize{\scriptsize}
\tablewidth{0pt}
\tablecaption{Derived Properties \label{tab:calc}}
\tablehead{
\colhead{Source}           & \colhead{$\ell$}      &
\colhead{$b$}         & \colhead{PDR\tablenotemark{a}} &
\colhead{$y^+$}   & \colhead{$\sigma \, y^+$}       & \colhead{$T_e$} & \colhead{$\sigma \, T_e$}  & \colhead{$T_{d}$} \\
\colhead{}           & \colhead{(deg.)}      &
\colhead{(deg.)}  & \colhead{}       &
   &      & \colhead{(K)} & \colhead{(K)} & \colhead{(K)}}
\startdata
E5 & 111.649 & 0.813 & out  & $<$0.149 & \nodata & \nodata & \nodata & 20.6\\
E4 & 111.627 & 0.813 & out  & $<$0.045 & \nodata & \nodata & \nodata & 20.6\\
E3.5 & 111.616 & 0.813 & out & $<$0.066 & \nodata & \nodata & \nodata & 20.5\\
E3 & 111.605 & 0.813 & out  & 0.091 & 0.008 & 10,550 & 180 & 19.3\\
E2 & 111.582 & 0.813 & on  & 0.083 & 0.002 & 8,330 & 50 & 21.8\\
E1 & 111.559 & 0.813 & in  & 0.092 & 0.003 & 8,250 & 60 & 22.8\\
C & 111.544 & 0.813 & in  & 0.086 & 0.002 & 7,890 & 50 & 23.9\\
W1 & 111.530 & 0.813 & in  & 0.088 & 0.001 & 7,520 & 20 & 26.4\\
W2 & 111.507 & 0.813 & on  & 0.091 & 0.003 & 9,440 & 60 & 25.8\\
SE & 111.603 & 0.762 & out & $<$0.063 & \nodata & \nodata & \nodata & 19.8
\enddata
\tablenotetext{a}{\,``out"\,...\,outside the PDR, ``on"\,...\,on the PDR, ``in"\,...\,inside the PDR.}
\tablecomments{All uncertainties are $\pm$1$\sigma$.}\vspace{-20pt}
\end{deluxetable*}

\subsection{Electron Temperature}
\label{sec:te}
The electron temperature, $T_e$, is an important physical parameter of \hii\ 
regions that can be used to assess the efficiency of cooling processes. 
There is debate in the literature
about the relationship between the electron temperature inside and
outside the \hii\ region PDR.  While previous observations of
\hii\ regions have found a fairly homogeneous spatial $T_e$ distribution
\citep[see][]{Roelfsema1992,Adler1996,Krabbe2002,Rubin2003}, an increase in scatter
has been observed at the edges of \ngc346 \citep{Oliveira2008}.  For
the well-studied Orion nebula, the results are inconclusive, with
optically-derived values of $T_e$ the same inside and outside the PDR
\citep{Mccall1979} or different findings for different species
\citep{Walter1994, Weilbacher2015}.  Recently, \citet{Wilson2015}
found a decrease in the radio-derived electron temperature with angular offset for Orion A.

We calculate the electron temperature assuming local thermal
equilibrium (LTE) for all RRL positions where a He line could be
detected, using
\begin{multline} \left( \frac{T_e^*}{\textnormal{K}} \right) = \left \{ 7103.3 \left( \frac{\nu _L}{\textnormal{GHz}}\right)^{1.1} \left[ \frac{T_C}{T_L(\textnormal{H}^+)} \right] \times \right. \\ \left. \left[ \frac{\Delta V (\textnormal{H}^+)}{\textnormal{km\,s}^{-1}} \right]^{-1} \times
\left[ 1+y^+ \right]^{-1} \right \}^{0.87},\label{eq:etemp}
\end{multline}
where $\nu_L = 8.9$\,GHz is the average frequency of our
Hn$\alpha$ recombination lines, $T_C$ is the continuum antenna
temperature, $T_L$ is the H line antenna temperature, $\Delta V
(\textnormal{H}^+)$ is the FWHM line width, and $y^+$ is the ionic
abundance ratio found by Eq.\,\ref{eq:heh} \citep[see][]{Quireza2006}. 
We derive errors in $T_e$ by propagating errors in $T_C$, $T_L$, 
$\Delta V(\textnormal{H}^+)$, and $y^+$ via Eq.\,\ref{eq:etemp}.
Here, the uncertainty of the continuum antenna
temperature, $\sigma T_C$, is due to random temperature
fluctuations found by taking the rms of an Off-source location in the map close
to \ngc7538. Uncertainties of all line
parameters are based on the Gaussian fits used to characterize the
line.

\begin{figure}
\figurenum{8}
\centering
\includegraphics[width=0.5\textwidth]{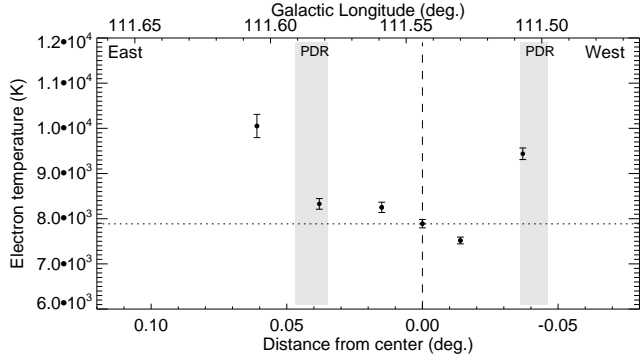}
\caption{LTE electron temperature, $T_e$, as a function of distance from
  the central location. The x-axis and meaning of the
  shaded regions are the same as in Figure~\ref{fig:IRpdr}. The dotted horizontal 
  line shows the average $T_e = 7890$\,K inside the PDR. The
  electron temperature remains relatively constant inside the PDR, but we detect
  significantly higher values of $T_e$ on and outside the PDRs. Error bars are
  $\pm$1$\sigma$. \label{fig:Etemp}}
\end{figure}

We show the derived electron temperature values as a function of
distance from the central location in Figure~\ref{fig:Etemp}. We find
an average temperature of $T_e = 7890 \pm 300$\K within the PDR of 
the region.  The peak of the radio
emission (position W1) has $T_e = 7520 \pm 20$\,K. Our E3 position outside 
the PDR is much hotter with $T_e = 10050 \pm 180$\,K, similar to the W2 position 
on the PDR ($T_e = 9440 \pm 60$\,K). The electron temperature at the E2 
position that is also located on the PDR, however, is closer to the average 
value inside with $T_e = 8330 \pm 50$\,K. It appears that $T_e$ remains roughly
constant within the PDR but may increase outside the PDR boundary. Previous work 
at the same observing frequencies reports average electron temperatures for \ngc7538 
of $8230 \pm 40$\K \citep{Quireza2006} and $8483 \pm 51$\K \citep{Balser2011} 
which are slightly higher than our results. This is likely due to uncertainties in the observed continuum antenna temperatures. \citet{Quireza2006} assigns a quality factor of ``B" to the continuum data of NGC\,7538, whereas \citet{Balser2011} gives a quality factor of only ``C" (``A" is highest quality). Additionally, observations at slightly different locations may result in differences in the calculated electron temperatures. \citet{Quireza2006} observed at ($\ell$,\,$b$) = (111.53\,\degree,\,0.82\,\degree) and \citet{Balser2011} observed at ($\ell$,\,$b$) = (111.525\,\degree,\,0.816\,\degree). These positions are offset from our W1 position by 25$''$ and 22$''$, respectively.

\subsection{Dust properties}
\label{sec:dust}
Higher dust temperatures outside the PDR may be correlated with
the locations along the PDR that have significant fractions of
escaping radiation. \citet{Anderson2012} and \citet{Anderson2015}
show that for the \hii\ region RCW\,120 the locations of dust
temperature enhancements outside the PDR are correlated
with ``holes'' in the PDR. These results indicate that radiation
can escape through these holes to heat the ambient medium. This
effect has yet to be investigated in detail for other \hii\ regions,
although it may be related to the extended radio continuum emission
around ultra-compact \hii\ regions observed by \citet{Kim2001}.

We use \textit{Herschel Space Observatory} data at 160, 250,
350, and 500\,\micron\ from the HOBYS key-time program
\citep{Motte2010} to derive dust temperature and column density maps
of \ngc7538. The HOBYS data of \ngc7538 were first shown in
\citet{Fallscheer2013}.  Here we use a similar method to create
these maps by fitting grey-body models pixel-by-pixel, after
regridding to the common spatial resolution of the
350\,\micron\ data \citep[see][]{Anderson2012}.  In the fits we
assume a dust emissivity index $\beta = 2$.  We subtract an offset
from all data at the position $\lb = (113.271\,\degree,
0.425\,\degree)$, as this location is relatively devoid of
emission. To convert to column density, we assume a dust-to-gas
ratio of 100, and an opacity $\kappa_\nu = 0.1 (\nu/1000
\ghz)^\beta$.

\begin{figure*}
\figurenum{9}
\centering
\begin{tabular}{rl}
\includegraphics[width=0.48\textwidth]{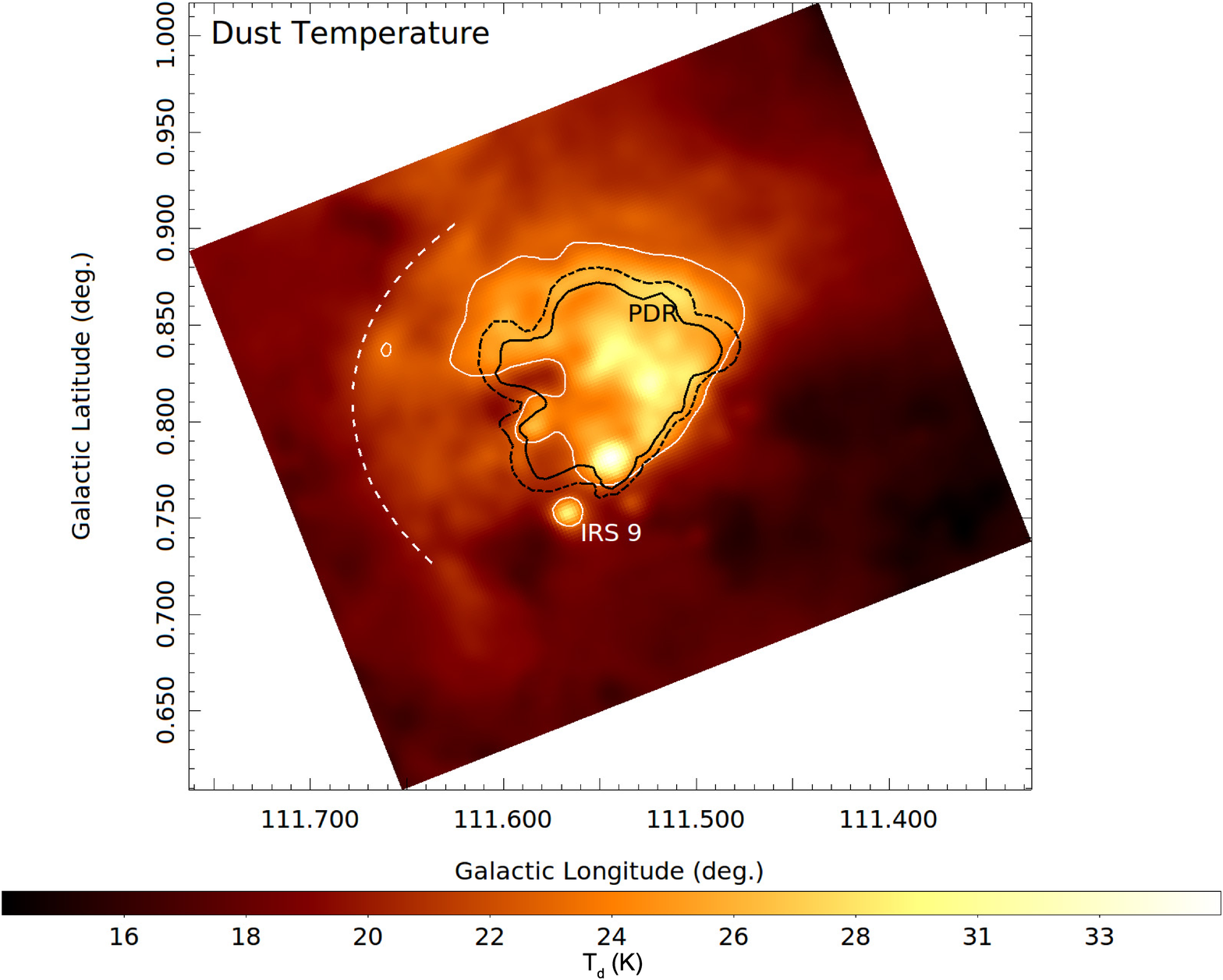}&
\includegraphics[width=0.48\textwidth]{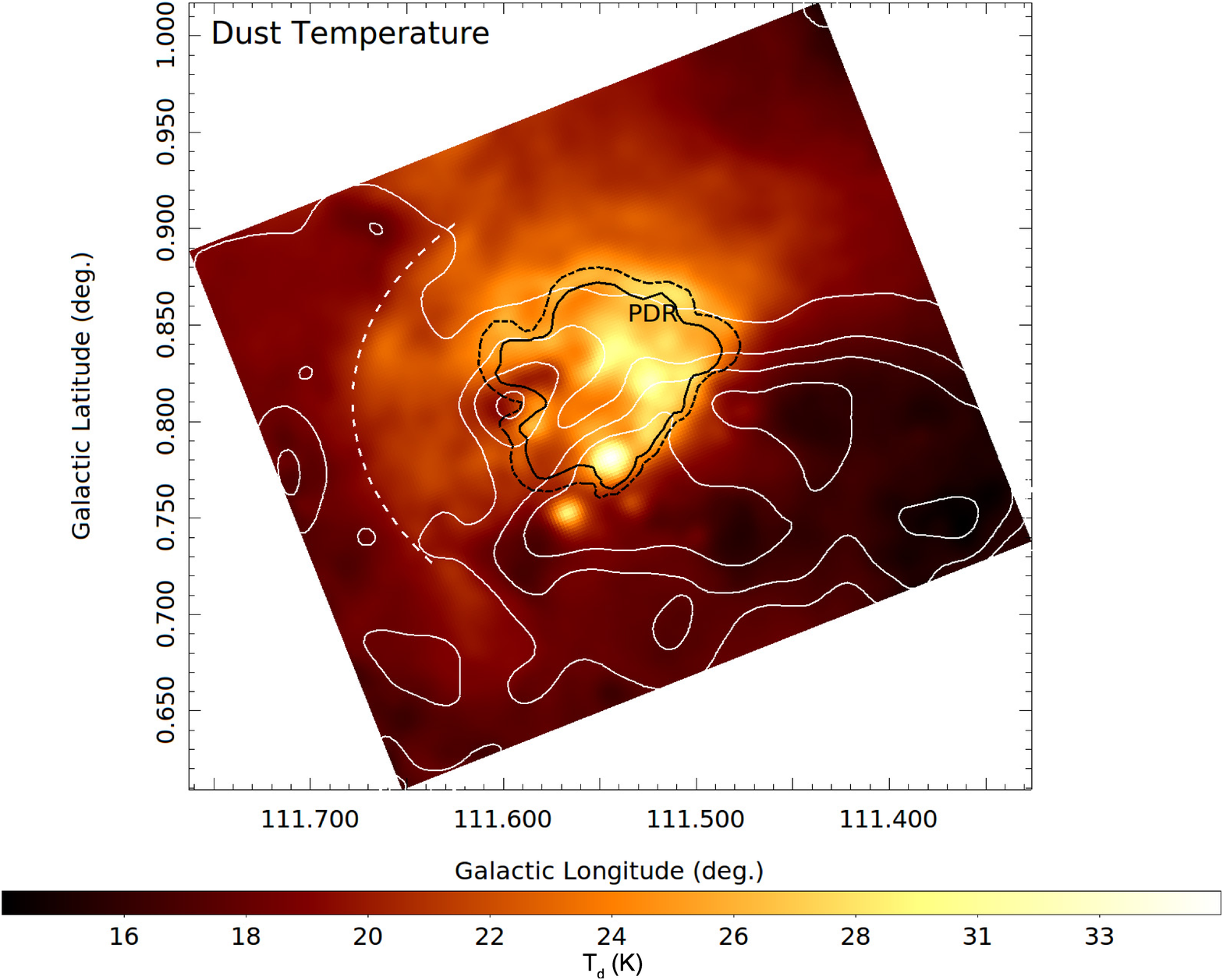}
\end{tabular}
\caption{Left: Dust temperature map of \ngc7538, derived from {\it
  Herschel} data \citep[see][]{Anderson2012}.  The solid white
  outline indicates the boundary of 23\,K dust temperature (hotter
  dust inside).  The inner/outer PDR boundaries are indicated by the
  solid/dashed black lines marked ``PDR". The second PDR is indicated
  by the dashed white line to the east.  Significant temperature
  enhancements are observed outside the PDR in the northern and
  eastern part of the region, possibly due to leakage through the PDR.
  These temperature enhancements to the east are confined to the
  second PDR boundary. The point-like enhancement to the south of the
  region is associated with the infrared source IRS\,9 and is not
  related to dust heating by \ngc7538 itself. Right: Same
  image as left, except that the white contours are of column
  density. Contour levels are at $3$, $6$, $10$, and $20 \times
  10^{21}$\,cm$^{-2}$.  Locations along the PDR of high column
  density are associated with low dust temperatures, and vice versa,
  showing how unimpeded radiation can heat the local ambient medium.
  \label{fig:heat}}
\end{figure*}

We show the dust temperature map of \ngc7538 in
Figure~\ref{fig:heat}. The contour in the left panel of
Figure~\ref{fig:heat} is at 23\K, as this temperature best
highlights the features of the dust temperature distribution near
the PDR. We see higher dust temperatures to the north and east of
the PDR compared to the west and south,
which may suggest that leaking radiation is heating the ambient
material. Within the second PDR to the east (defined in
Figure~\ref{fig:IRbeam}) we detect higher dust temperatures of
$>$22\K compared to $\sim$20\K outside.  This indicates that the
second PDR is an additional barrier to the propagation of photons
from the region further into the ISM.  The point-like temperature
enhancement to the south of the region is associated with the
infrared source IRS\,9 and is not related to dust heating by
\ngc7538 itself.

The contours in the right panel of Figure~\ref{fig:heat} are of
column density.  The dust temperature is higher in directions where
the column density along the PDR is lower, namely toward the north
and east.  There is a small region of high column density on the
eastern PDR.  This region is presumably leading to the two ``lobes''
of radio emission seen in this direction (Figure~\ref{fig:Loc}).
Toward the south and west there is high column density material
along the PDR that is spatially correlated with lower dust
temperatures.

\section{Discussion}
Studies of external galaxies suggest that $30-70$\,\% of the emitted
hydrogen-ionizing photons escape from \hii\ regions into the ISM
\citep[e.g.][]{Oey1997,Zurita2002,Giammanco2005,Pellegrini2012}. In our 
Galaxy, however, information on radiation leaking from \hii\ regions is 
sparse. Furthermore, Galactic studies are often contaminated by confusion 
since we reside in the Galactic plane \citep[see][]{Anderson2015b}. Here we 
use the Outer Galaxy \hii\ region NGC\,7538 as a case study since there is 
less confusion along the line of sight to this source.

\begin{figure}
\figurenum{10}
\centering
\vspace{-15pt}
\begin{tabular}{c}
\vspace{-20pt} \includegraphics[width=.47\textwidth]{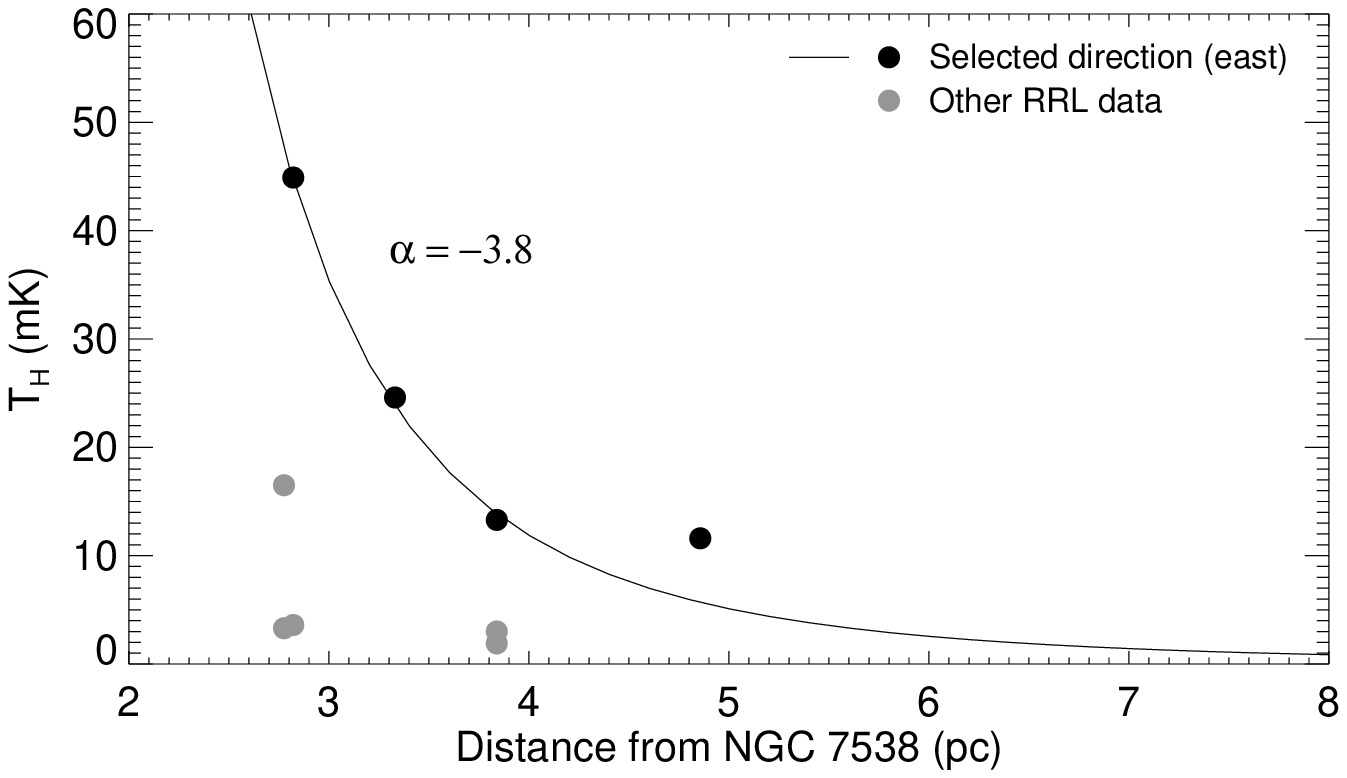} \\
\includegraphics[width=.47\textwidth]{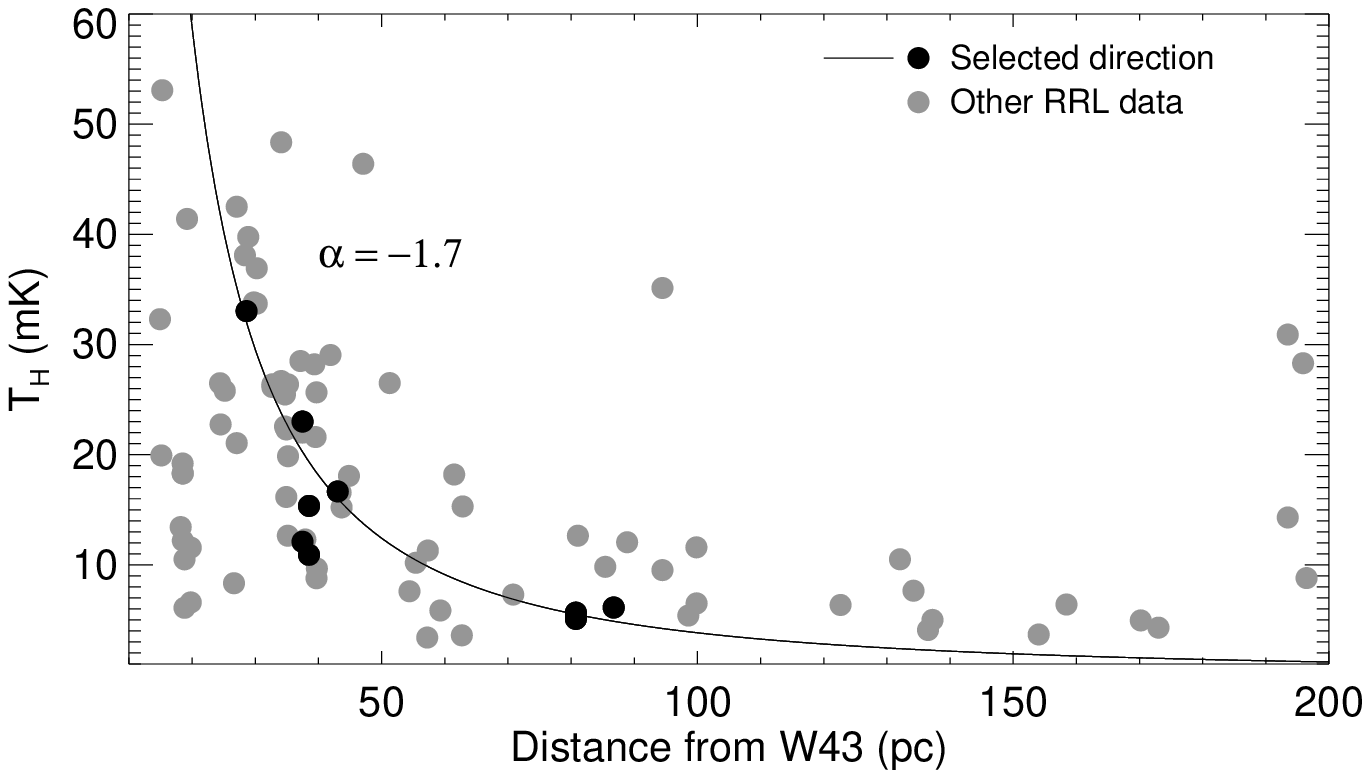}
\end{tabular}
\caption{The RRL intensity outside the PDR of \ngc7538 (top) and W43
  (bottom) as a function of distance from the nominal center.
  Due to insufficient data points in the other directions, 
  we only fit data from the east direction for \ngc7538. Due to the asymmetrical
  nature of W43, we must fit the data separately for different directions
  from the region. The shown fit for W43 represents a 10\,\degree sector centered at 295\,\degree from Galactic north. The emission from 
  \ngc7538 decreases more rapidly than that from W43 as indicated by the power-law
  index $\alpha$ in the figures. \label{fig:powerlaw}}
\end{figure}

The detection of radio continuum and RRL emission outside
the PDR of \ngc7538 hints that the ionizing photons from \hii\ regions
could be responsible for some of the diffuse RRL
emission detected in the plane of the Galaxy \citep{Anderson2011,Anderson2015b}.  
The intensity of this emission outside the PDR, however, decreases 
rapidly with increasing distance.
To quantify this decrease, we fit a power law of the form 
$T_L [\textnormal{mK}] = T_{L,0} d^{\alpha}$ 
to the H RRL intensity outside the PDR. Here, the model parameter 
$T_{L, 0}$ is the antenna temperature value at a distance of 1\,pc from the 
center of the region at ($\ell$,\,$b$) = (111.544\,\degree,\,0.813\,\degree), 
$d$ is the distance in pc from the center, and $\alpha$ is the
power law index.
Since the antenna temperature on the PDR varies with direction, 
we have to fit each direction separately. We only fit RRL intensities
towards the east of the region due to a lack of data points in the 
other directions. We find $\alpha = -3.8$, indicating a very steep
decrease with distance. These results are
shown in the top panel of Figure~\ref{fig:powerlaw}.

We repeat this analysis for the complex W43 region
(bottom panel of Figure~\ref{fig:powerlaw}) using RRL data from \citet{Anderson2015b}
for which the emission has been identified as ``diffuse'' (i.e. not
from within the PDR of a discrete \hii\ region).  These data were
taken and reduced using the same methodology employed here and
so are directly comparable to our data for \ngc7538.  We use all RRL
detections within 2\,\degree of W43 that have velocities between 80 and
120\,\kms\ and assume the maser parallax distance for W43 of 5.5\,kpc
\citep{Zhang2014}. We split our analysis into sectors of 10\,\degree 
opening angle around W43 and fit each separately. This allows us to 
detect spatial asymmetries of the diffuse emission around W43 similar 
to the analysis of \ngc7538 above. The power law fits range from 
$\alpha = -0.26$ to $\alpha = -3.01$, with an average of $\alpha = -0.54$, 
indicating that the decrease in diffuse RRL intensity for W43 is 
significantly less steep than that for \ngc7538. A representative power
law fit (the 10\,\degree sector centered at 295\,\degree from
Galactic north) for W43 is shown in Figure~\ref{fig:powerlaw} for illustration 
purposes.

What would lead to the difference in power law indices?  W43 is much
more luminous than \ngc7538.  \citet{Smith1978} list a flux density of
86.5\,\jy\ for W43 at 5\,\ghz. Given their respective Heliocentric
distances, and assuming that both regions are optically thin, this
implies that W43 is $\sim$20 times more luminous than \ngc7538.
\citet{Zurita2000} show that in a model suggested by 
\citet{Beckman2000}, leaking radiation from very luminous \hii\ regions 
is sufficient to ionize the diffuse gas. The model, based on a change in 
slope in H$\alpha$ luminosity functions of \hii\ regions, assumes that high
luminosity \hii\ regions are primarily density-bounded. This is consistent 
with their data on a sample of six spiral galaxies and would allow a 
large number of ionizing photons to escape into the surrounding medium.
If the extended emission around W43 is due to W43 itself, the shallow
power law may imply that this is the case here. As a result, a 
significant fraction of the leaked radiation would be able to ionize the 
diffuse gas. Along the W43 sight line, 
however, it is difficult to disentangle the possible contribution from the 
numerous \hii\ regions. We therefore cannot be sure that all the diffuse
radiation is due to W43 itself. Another model, proposed by
\citet{Anantharamaiah1986}, suggests that the extended emission could be due 
to low-density envelopes of size $\sim$100\,pc around individual compact \hii\
regions. Due to the high population of \hii\ regions around W43, our line of 
sight could cross many of these envelopes. \citet{Roshi2001} argue that this 
model is in good agreement with RRL observations at 327\,MHz in the Galactic 
plane.

The steep power law of \ngc7538 shows the difficulty in detecting radio
emission far from such compact \hii\ regions.  We see from
Figure~\ref{fig:powerlaw} that 5\,pc from the center the fitted RRL
intensity is just $\sim$5\,\mK.  Furthermore, there is clearly a second
PDR seen toward the east, visible at 12\,\micron\ in
Figure~\ref{fig:IRbeam}.  The zone interior to this second PDR has a
higher dust temperature with a sharp boundary (see Figure~\ref{fig:heat}).
This implies that in this direction the radiation is leaking
through the primary PDR.  This radiation, however, is being 
further attenuated by the secondary PDR and is not truly escaping the region.

Our results cast doubt on whether compact \hii\ regions like
\ngc7538 can be responsible for the diffuse ionized gas detected in the 
inner Galaxy. Since this is a case study of only one such region, however, the broader implications are unclear. While giant \hii\ region complexes such as W43 seem to leak a significant fraction of their ionizing radiation into the ISM, more study is required on individual regions. Larger \hii\ regions in the Outer Galaxy such as W3/W4/W5 may be suitable targets for future observations as there is little confusion along the line of sight.
\section{Summary}
We observed the \hii\ region \ngc7538 with the NRAO Green Bank
Telescope (GBT) in $\sim$8.7 GHz continuum and recombination line
emission to investigate the escape of radiation through the PDR into
the local interstellar medium.  This leaked radiation from \hii\ regions
could be an important source of photons that maintain the ionization
of the WIM.

Using the radio continuum intensity, we estimate the leaked radiation
fraction from \ngc7538. We first visually define the PDR using WISE 12\,$\mu$m
infrared images and sum the pixel values of the radio 
continuum map inside and outside the PDR to find the total intensity 
of \ngc7538. We calculate a total intensity of 17.3\,Jy inside the region which
would roughly be expected from a single O5V
star. Using the ratio of the outside to total radio continuum intensity, 
we compute the percentage of leaked emission, $f_R$, to be 
$f_R = 15 \pm 5$\,\% in the plane of the sky.  We also use H$\alpha$ data
from IPHAS to repeat the calculation and find $f_{H \alpha} = 31 \pm
10$\,\%. Due to the three-dimensional geometry of the region, these numbers 
represent lower limits on the total leaked emission.

The leaking radiation is not found everywhere around the region,
but rather is mostly in the north and east.  These are also the
directions where we identify dust temperature enhancements, and
locations along the PDR of decreased column density.  This shows
how, due to a non-uniform PDR, radiation can escape the region in
some directions to heat the ambient medium while it is confined in
other directions.

RRL measurements of \ngc7538 are consistent with a decrease 
in the ionic abundance number
ratio $y^+$ with increasing distance outside the PDR and an increased 
carbon line emission in the PDR, suggesting a
softening of the radiation field. We compute an
average LTE electron temperature of $7890 \pm 300$\,K inside the
region, which is marginally lower than that derived in previous work
\citep{Balser1999,Quireza2006}.

The RRL intensity from \ngc7538 decreases rapidly outside the PDR.
This decrease is much steeper than that seen for the high mass star
formation complex W43. Furthermore, the existence of an additional PDR
boundary further from \ngc7538 implies that the ``leaked'' emission is
not able to traverse large distances.  This seems to imply that
maintaining the WIM might rather be caused by giant \hii\ region
complexes regions such as W43 than ``normal'' compact \hii\ regions
like \ngc7538.

\acknowledgements
We thank West Virginia University for its financial
support of GBT operations, which enabled the observations for this
project. \nraoblurb

This paper makes use of data obtained as part of the INT
Photometric H$\alpha$ Survey of the Northern Galactic Plane (IPHAS,
www.iphas.org) carried out at the Isaac Newton Telescope (INT). The
INT is operated on the island of La Palma by the Isaac Newton Group in
the Spanish Observatorio del Roque de los Muchachos of the Instituto
de Astrofisica de Canarias. All IPHAS data are processed by the
Cambridge Astronomical Survey Unit, at the Institute of Astronomy in
Cambridge. The bandmerged DR2 catalogue was assembled at the Centre
for Astrophysics Research, University of Hertfordshire, supported by
STFC grant ST/J001333/1.

\facility{Green Bank Telescope.}
\software{TMBIDL \citep{Bania2014}.}

\bibliographystyle{aasjournal}
\bibliography{HII}

\end{document}